%% file: main.tex
\documentclass[fleqn,10pt]{wlscirep}
\usepackage[utf8]{inputenc}
\usepackage[T1]{fontenc}
\usepackage{tcolorbox}
\usepackage{siunitx}
\usepackage{xcolor}
\usepackage{fvextra}

\usepackage{csquotes}
\usepackage[numbers]{natbib}
\usepackage{threeparttable}
\definecolor{mBlue}{HTML}{407193} 
\definecolor{mOrange}{HTML}{ca8243} 
\title{Closing the Loop to Discover Psychological Theories with an Automated Cognitive Scientist}

\author[1, +, *]{Akshay K. Jagadish}
\author[2, +]{Younes Strittmatter}
\author[3]{Nori Jacoby}
\author[4]{George Kachergis}
\author[5]{Eric Schulz}
\author[2, 6]{Nathaniel Daw}
\author[2, $\psi$, *]{Suyog H. Chandramouli}
\author[1, 2, $\psi$]{Thomas L. Griffiths}
\affil[1]{Princeton AI Lab, Princeton University}
\affil[2]{Department of Psychology, Princeton University}
\affil[3]{Department of Psychology, Cornell University}
\affil[4]{Department of Psychology, Stanford University}
\affil[5]{Institute for Human-Centered AI, Helmholtz Munich}
\affil[6]{Princeton Neuroscience Institute, Princeton University}
\affil[$\psi$]{indicates equal senior author contribution, listed in alphabetical order}
\affil[+]{indicates equal first author contribution}
\affil[*]{corresponding authors: akshaykjagadish@gmail.com, suyoghc@gmail.com}


\begin{abstract}
Across the sciences, autonomous systems are increasingly being used in closed-loop discovery, proposing new theories and designing and running experiments to test them. This approach is yet to be applied in the field of cognitive science, where the central bottleneck is theory-building: the creative step of turning the accumulated failures of existing models into better ones. Theory generation has remained manual even as data collection, modeling, and experiment design have been automated. We present the \textsc{Auto}mated \textsc{Cog}nitive Scientist (\textsc{AutoCog}), a fully autonomous agentic-AI system that closes this loop. Large-language-model agents advocate competing theories, each expressed as an executable cognitive model, design experiments that best discriminate them, collect behavioral data from participants recruited online, score theories against collected data based on their generative performance, diagnose why they fail, and synthesize a better successor. Repeating this cycle allows them to search the space of theories, models, and experiments. In the domain of decision-making, \textsc{AutoCog} recovered known decision-making strategies from simulated behavior, including unconventional ones, showing that its discoveries are ultimately driven by the data rather than strictly bound by the priors of the underlying language models. When run with human participants, it produced theories that outperformed the established theories it was seeded with and generalized to held-out studies in two different experimental settings. It also surfaced a novel theory of multi-cue decision-making in which choices show diminishing sensitivity to feature values. The distinctive predictions of this theory were  confirmed in a preregistered study with new participants. \textsc{AutoCog} demonstrates how  an automated discovery system can be used to turn cognitive theory-building into an explicit, executable, and cumulative science.

\end{abstract}
\begin{document}

\flushbottom
\maketitle

\thispagestyle{empty}







\section*{Introduction}


Scientific discovery is increasingly carried out by autonomous systems \citep{musslickAutomatingPracticeScience2025, wangScientificDiscoveryAge2023,szymanski2023autonomous,lu2026towards, swanson2025virtual}. Such systems are most valuable when they execute the entire discovery cycle rather than automating any single stage: proposing experiments, collecting evidence, comparing competing explanations, and using the result to plan new experiments or revise theories. In chemistry and materials science, self-driving laboratories design and run their own experiments \citep{abolhasaniRiseSelfdrivingLabs2023, szymanski2023autonomous}; in computer science, AlphaEvolve has discovered faster algorithms \citep{novikov2025alphaevolve}; and in mathematics, AI has helped resolve long-standing open problems \citep{feng2026semiautonomous}. 
Psychology is well-positioned to automate the full scientific discovery cycle \cite{musslick2024closed}. 
Like other empirical sciences, it studies a phenomenon of nature, i.e. human behavior, yet does not require wet lab or field experiments;
many psychology experiments can be run online with human participants and analyzed in code, so the entire loop can be automated end to end \cite{musslick2024autora}. 
In recent years, several parts of this discovery cycle have accelerated:
online platforms and large-scale experiments have expanded behavioral data collection \citep{hartshorneCriticalPeriodSecond2018a, jones2017big}; cognitive models can be discovered from  behavioral datasets \citep{peterson2021using, rmusGeneratingComputationalCognitive2025, castro2025discovering, xie2026think, kasenberg2026ai}; and optimal experimental design can select informative experiments that discriminate between candidate models \citep{cavagnaroAdaptiveDesignOptimization2010, eltetHo2026atlas}. 

Yet, closing the loop on human behavior remains a largely unmet challenge \cite{jagadishCanWeAutomatize2026}.
The difficulty is more conceptual: the claims are not trivially verifiable, as in other natural sciences such as mathematics, and a theory of the mind cannot be compiled, solved, or synthesized \citep{guestMartinForceTheory2021, vanRooijBaggioTheoryBeforeTest2021, eronenTheoryCrisisPsychology2021}, but instead must navigate an open conceptual space to render opaque psychological processes mechanistically meaningful.
Therein, the central bottleneck to closing the loop has never been merely data collection or modeling, but rather the creative, historically human art of using the accumulated empirical failures of existing models to imagine better ones \citep{eronenTheoryCrisisPsychology2021, muthukrishnaProblemTheory2019}.

\begin{figure*}[t!]
\centering
\includegraphics[width=1.\linewidth]{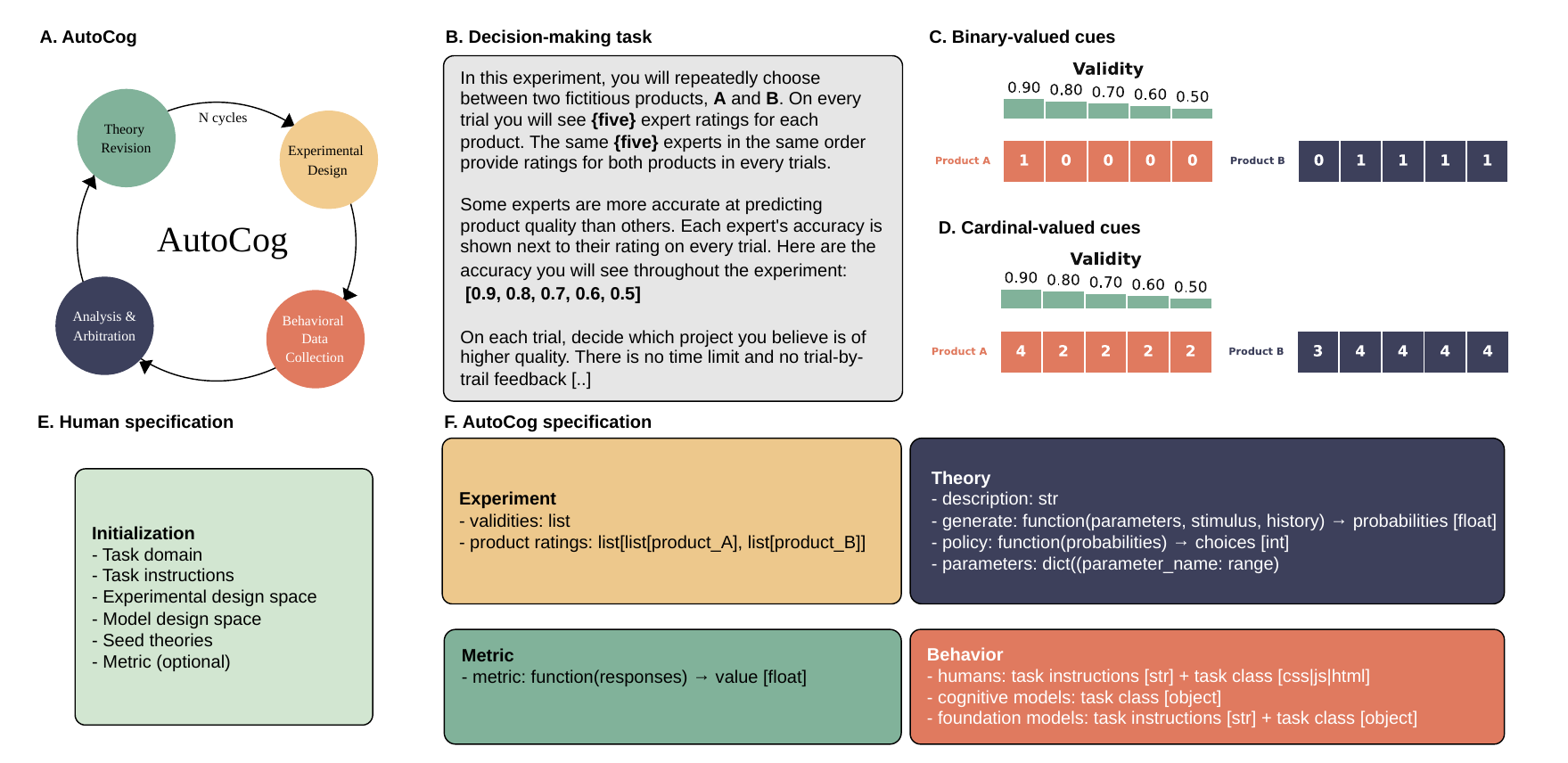}
\caption{\textbf{The \textsc{AutoCog} closed-loop discovery system.}
  \textbf{(A)}~\textsc{AutoCog} runs as a closed discovery loop iterated over $N$ cycles:
  \emph{Experimental Design} $\rightarrow$ \emph{Behavioral Data Collection} $\rightarrow$ \emph{Analysis \& Arbitration}  $\rightarrow$ \emph{Theory Revision} , after which a revised theory re-enters the
  loop.
  \textbf{(B)}~The task framing for the decision-making task used throughout.
  \textbf{(C)}~A \emph{binary-valued} cue instantiation, in which each
  expert's rating is $0$ or $1$.
  \textbf{(D)}~A \emph{cardinal-valued} instantiation, in which expert ratings are non-negative
  integers between $0$ and $r_{max}$, displayed as horizontal filled bars annotated with their numeric value (e.g. ``4/5'').
  \textbf{(E)}~The human-supplied \emph{initialization}: task domain, task
  instructions, experiment design space, model design space, seed theories, and an
  optional metric.
  \textbf{(F)}~The corresponding \textsc{AutoCog} specification---the typed
  interfaces the system operates over: an \emph{Experiment} (validities and
  per-product ratings), a \emph{Theory} (a \texttt{generate} function mapping
  parameters/stimulus/history to choice probabilities, a \texttt{policy}, and
  parameter ranges), a \emph{Metric} (responses $\rightarrow$ scalar), and a
  \emph{Behavior} adapter that renders the same task for human participants,
  cognitive models, and foundation models.}
\label{fig:overview}
\end{figure*}

 Here we present \textsc{AutoCog} (\textsc{Auto}mated \textsc{Cog}nitive scientist), a closed-loop system for discovery of psychological theories from human behavior. This system uses large language model (LLM) agents to advocate for competing theories expressed as executable models, design experiments that discriminate between them, run them on human participants, evaluate the resulting behavior, and synthesize a successor theory from the resulting record; see Figure~\ref{fig:overview}A. 
  We applied \textsc{AutoCog} to decision-making, given its rich history and its mature theories supplemented by concrete cognitive models \citep{gigerenzer1999simple, gigerenzer2011heuristic, tversky1974judgment, kahneman2013prospect}. Specifically, we considered multi-attribute decision-making \citep{hilbig2014generalized}, where people choose between options described by several expert ratings
  of differing reliability, and evaluated the system in two settings; see Figure~\ref{fig:overview}B-D. First, we tested whether the loop could recover the ground-truth strategy behind simulated behavior, both for canonical decision heuristics and for
  intentionally unconventional strategies. Second, we closed the loop directly with human participants recruited online, where LLMs take over the expert-driven step of proposing revised theories while a generative simulation-based comparison against the collected data determines which candidates are retained, revised, or replaced. In both settings, \textsc{AutoCog} moved beyond the theories with which it was seeded to surface accounts that predicted held-out behavior better than either seed. Beyond predicting behavior more accurately, the loop surfaced a previously unreported regularity in how people choose: choice shows diminishing sensitivity to cue value, a prediction we confirmed prospectively in a preregistered study with new participants. To our knowledge, \textsc{AutoCog} is the first autonomous closed-loop system to discover theories of human cognition directly from human behavior.

\section*{Automated cognitive scientist (\textsc{AutoCog})}


Theory building in psychology follows a recognizable cycle \cite{simon2012models}. 
Consider a cognitive scientist theorizing about how the human mind solves a particular task. 
They seldom begin from a blank slate; they hold competing theories of the process and want to know which is closer to the truth \cite{chamberlin1890method, robertsPashler2000}. Rather than collecting data from an arbitrary experiment, they design one where the two competing accounts disagree \cite{platt1964strong, meehl1978theoretical}. They administer that experiment to human participants, compare each theory's predictions against what the participants actually did, and ask not only which account fared better but why the other failed \cite{experimentology2025}. That diagnosis then feeds into constructing a better theory and the cycle repeats. 

\textsc{AutoCog} is a fully autonomous agentic system that runs this cycle end-to-end using agents based on LLMs.\footnote{The default LLM used throughout this work was \texttt{gemini-3.1-pro-preview} unless otherwise mentioned}
It begins with the researcher specifying the starting conditions (Figure \ref{fig:overview}E): the task domain and instructions, an experiment design space that bounds the possible experiments, a model design space that fixes its inputs and outputs, and two seed theories to begin the search.
\textsc{AutoCog}  then enters a cyclic process with four key stages (see Figure \ref{fig:overview}F):

\begin{itemize}

     \item[1.]  \textbf{Experimental design:} An experiment where the two incumbent theories make divergent predictions is proposed by an LLM-agent following the principle of adversarial experimental design \cite{chandramouli2026automated}. In conjunction with the experiment, an LLM-agent also proposes a metric that quantifies the predicted advantage of one theory over the other \cite{li2024critical}.

    \item[2.]  \textbf{Behavioral data collection:}  Behavioral responses for the committed experiment are programmatically \cite{sweetbean, sweetpea} collected from one of three sources: humans through online behavioral study \cite{palan2018prolific}, foundation models as an in-silico stand-in \cite{binzFoundationModelPredict2025}, or cognitive models \cite{polk2002cognitive, l2008bayesian}. 
    
    \item[3.] \textbf{Analysis \& Arbitration}: The generative behavior of the incumbent theories, specifically its associated cognitive model, is compared with the observed behavior using pre-registered metrics  \citep{valentin2021bayesian, aushev2023online, l2008bayesian}. The resulting evidence goes through an LLM-agent-based arbitration procedure, returning an interpretation of the current findings and a recommendation to revise the current theories \cite{zhuge2024agent, zhu2026demystifying}.

     \item[4.] \textbf{Theory revision}: The weaker of the two theories is then revised based on the recommendation by a separate LLM-agent, either re-synthesizing a cognitive model associated with a theory or regenerating a theory from scratch using program synthesis \cite{rmusGeneratingComputationalCognitive2025, chandramouli2026automated, xie2026think}, before it returns to stage one to begin a new cycle.

\end{itemize}

In each stage, an LLM agent is prompted in natural language (see SI for exact prompts), with the relevant artifacts placed in its context, and asked to return the appropriate output for that stage: a set of experimental stimuli and a metric (stage 1), an arbitration verdict (stage 3), or a theory expressed as an executable Python program (stage 4). Every output is verified programmatically before it is accepted with the LLM agent prompted iteratively with feedback until a valid, verified output is produced or an attempt budget is exhausted; see Methods and SI for full details. 

Four principles run through \textsc{AutoCog}'s discovery loop and organize how its stages fit together. It treats discovery as an open-ended search over theories and experiments rather than selecting a fixed set of candidates \cite{stanley2015greatness}. It relies on generative behavior to compare theories, rather than fitting models to observed behavior \cite{binzFoundationModelPredict2025}, as it is faster and less likely to overfit \cite{pitt2002good, roberts2000persuasive, xie2025centaur, namazova2025not}. It is self-verifying: every proposal, whether an experiment or a theory, is verified against its own predictions before it observes any actual data \cite{li2024critical}. And it favors theories that integrate across tasks and populations, since a candidate theory progresses between cycles only by capturing all the behavior collected so far rather than being tuned to one dataset \cite{almaatouq2024beyond, yarkoni2022generalizability, newell2012you}. Together, this makes \textsc{AutoCog}\footnote{The version of \textsc{AutoCog} presented in this work is only one possible instantiation; we evaluated each of its design decisions against alternative options (see the Controls section in Methods).} a discovery engine that is open-ended, robust, scalable, and unifying \cite{newell1994unified}. 

\begin{figure}[!t]
    \centering
    \includegraphics[width=1.02\linewidth]{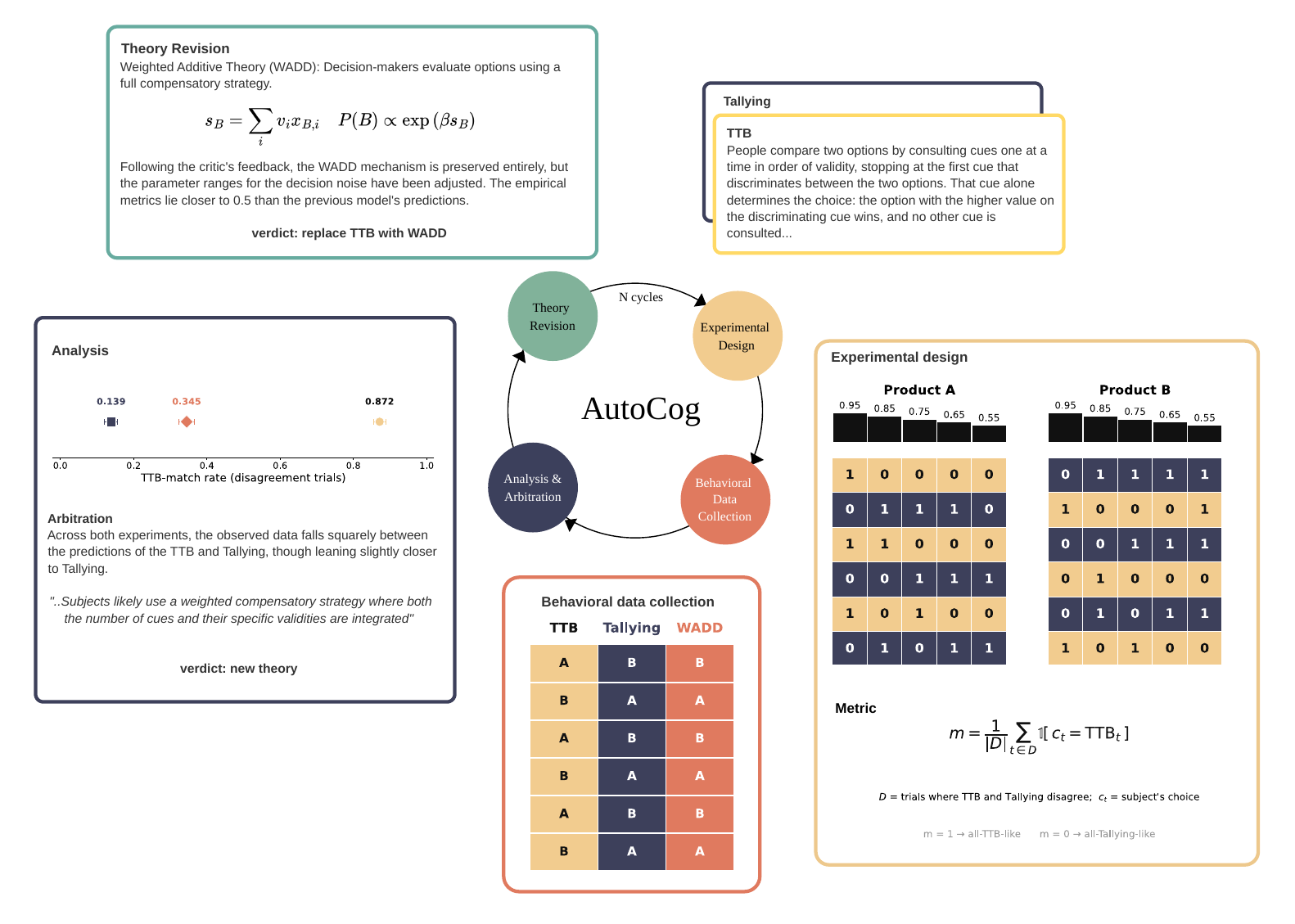}
    \caption{\textbf{Illustrative case-study of \textsc{AutoCog} on synthetic WADD-generated responses}
  A single discovery cycle (center) where \textsc{AutoCog} is tasked to recover WADD while being \emph{Seeded} with TTB and Tallying (top-right). 
  \emph{Experimental design} (right): the system
  constructs an experiment where the option pairs (cue matrices for Option~A and Option~B over five validity-ordered cues) are chosen to make the seed theories disagree.
  \emph{Behavioral response} (bottom): on each designed trial, the
  deterministic choices predicted by TTB, Tallying, and WADD.
  \emph{Metric} (right): the discriminating statistic on which TTB and Tallying make opposing choices.
  \emph{Analysis} (left): on the TTB-match rate over disagreement trials, the
  simulated participants fall between the TTB prediction and the Tallying prediction, nearer Tallying.
  \emph{Theory Revision} (top): the \emph{Arbitrator}, observing that the data fall squarely between TTB and Tallying, concludes that participants integrate both the \emph{number} of supporting cues and their \emph{validities} and issues a \emph{new-theory} verdict; the \emph{Theory Reviser} proposes WADD replacing TTB.}
    \label{fig:illustrativecasestudy}
\end{figure}

\section*{Results}

In the following sections, we demonstrate how \textsc{AutoCog} can recover not only known theories but also surface theories of decision making that predict behavior beyond the experiments and population-samples that produced them.

\subsection*{Closed loop recovery of known decision-making strategies}

\begin{figure}[t!]
    \centering
    \includegraphics[width=1\linewidth]{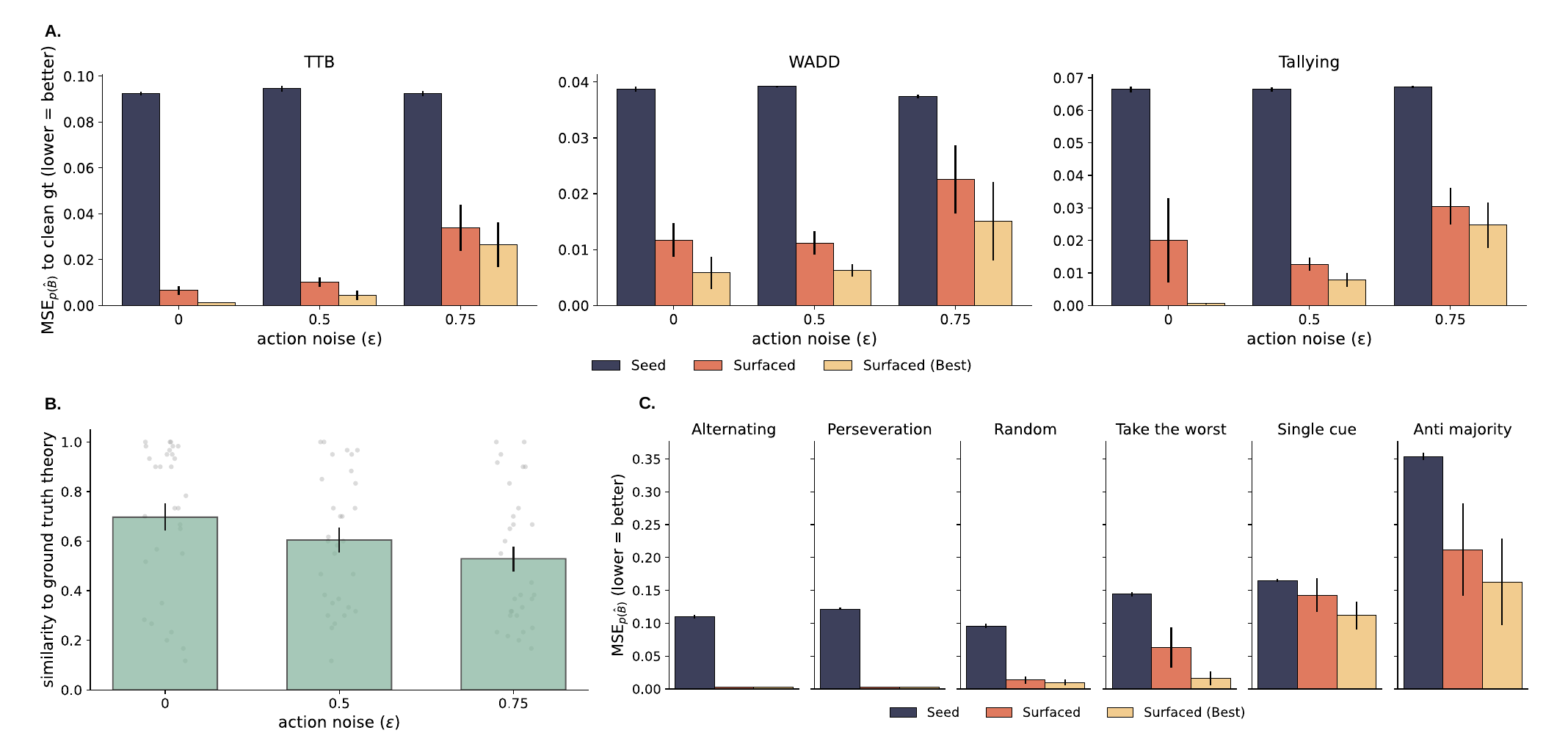}
    \caption{\textbf{Recovery of generating theories under increasing action
  noise.}   \textbf{(A)}~For each of the canonical decision-making strategies (TTB, WADD, Tallying), mean squared error of proportion of times option (B) is chosen for every unique option-pair $\hat{p}(B)$ against the \emph{noiseless} ground truth (lower $=$ better recovery), at action-noise levels $\varepsilon \in \{0, 0.5, 0.75\}$, which follows an$\epsilon$-greedy rule: each response was replaced by a uniform $50/50$ coin with probability $\epsilon$. Bars compare the \emph{Seed} theory   (indigo), the theory \emph{Surfaced} by \textsc{AutoCog} (peach), and the  \emph{best} surfaced theory (gold). 
  \textbf{(B)}~Mechanism similarity between each surfaced theory and the true
  generating theory (LLM-judge score, $0$--$1$) as a function of action noise;  gray points are individual theories and the black line is the pooled   mean~$\pm$~SEM.
  \textbf{(C)}~Recovery performance for six
  \emph{non-canonical} generating strategies (Alternating, Perseveration,
  Random, Take-the-worst, Single-cue, Anti-majority) at $\varepsilon=0$. Error bars are SEM across runs.}
  \label{fig:recovery}
\end{figure}

We first studied whether \textsc{AutoCog} can successfully recover a known cognitive theory when seeded with two other distinct theories. 
As reference theories, we chose existing accounts of human decision-making  \cite{tversky1974judgment,gigerenzer1996reasoning}, in particular Take-the-Best (TTB), Tallying, and Weighted-Additive (WADD). 
The simulations were conducted in a multi-attribute decision-making task with two options and binary-valued cues, as shown in Figure \ref{fig:overview}B.
We ran five cycles of \textsc{AutoCog} with behavioral data simulated from a held-out decision-making strategy, and the process was repeated for five runs each; see Methods for details. We provide an illustrative case-study in Figure \ref{fig:recovery} to make the different stages of \textsc{AutoCog} more concrete.

To estimate how well \textsc{AutoCog} recovers the ground-truth theory, we simulate the cognitive models associated with the true, surfaced, and seed theories on a held-out decision-making study~\cite{hilbig2014generalized}. For each model, we compute $\hat{p}(B)$, the proportion of times that option B is chosen, for every unique option-pair. We then quantify the alignment between two theories on the mean squared error ($MSE_{\hat{p}(B)}$) computed over the vector of $\hat{p}(B)$ values across all unique option-pairs. In the left-most bar for each theory in Figure \ref{fig:recovery}A, we report $MSE_{\hat{p}(B)}$ between the ground-truth and the two surfaced theories (peach), the best surfaced theory (gold), and the seed theories (indigo) under the noise-free condition.
We found that the theories surfaced by \textsc{AutoCog} ($M=0.013, SEM=0.004$) were much closer to the ground truth than the seed theories ($M=0.066, SEM=0.006$) and if the best of the two surfaced theories ($M=0.0025, SEM=0.0011$) are chosen, they are indistinguishable from the ground truth ($M=0.0006$; $p=0.10$, paired $t$-test), indicating that \textsc{AutoCog} can successfully recover ground truth theories in the absence of noise.
However, in practice, human responses tend to be noisy. We therefore tested performance across different noise levels by replacing each action with a random response with probability $\epsilon$ (see Methods). Specifically, we considered two noise-levels: 0.5 and 0.75, which roughly translates to picking an option different from the model in about 25\% or 37.5\% of the trials, respectively. We found that the ground-truth recovery performance worsens with noise, with \textsc{AutoCog}'s aggregate performance (in $MSE_{\hat{p}(B)}$) going from $0.011$ $(SEM=0.001)$ for $\epsilon=$ 0.5 and $0.029$ $(SEM=0.004)$ for $\epsilon=$ 0.75; see Figure \ref{fig:recovery}A.
In addition, we also used the LLM-as-judge framework \cite{gu2026survey, zheng2023judging} to assess whether \textsc{AutoCog} surfaced theories share the same underlying mechanism as ground-truth theory (see Methods for details). Pooled across canonical theories for the three different noise-levels, we found that the similarity of \textsc{AutoCog} surfaced theory to the ground-truth mechanism is highest under the noiseless condition ($M=0.697, SEM=0.055$) and slightly worse when the observed data are noisy: $\epsilon=$ 0.5 ($M=0.604$, $SEM=0.049$) and $\epsilon=$ 0.75 ($M=0.529$, $SEM=0.05$); see Figure \ref{fig:recovery}B. We provide an illustration of how the structure of surfaced theory changes with noise: from recovering a mechanism similar to ground truth to models with high lapse rate and eventually fully random-choice model; see SI.

As existing theories of human decision-making are present in the training data given to LLMs, we also tested how well \textsc{AutoCog} can recover non-canonical theories of human decision-making and a few spurious but well-observed behavioral policies \cite{miller2019habits}. Specifically, we considered six non-canonical strategies:  alternating, perseveration, random, take-the-worst, single cue, and anti-majority; see the baseline theories section in Methods for details. 
We found that \textsc{AutoCog} can successfully surface theories that recover alternating ($M=0.003, SEM=0.0004$), perseveration ($M=0.005, SEM=0.002$), random ($M=0.014, SEM=0.005$) and take the worst ($M=0.065, SEM=0.032$) strategies in five cycles. On manual inspection, we found that it was able to recover the exact underlying mechanism for these theories in at least two out of the three runs. For more unconventional theories, such as single-cue ($M=0.1412, SEM=0.0264$) and anti-majority ($M=0.2142, SEM=0.0689$), \textsc{AutoCog} can still surface theories that are behaviorally aligned to the ground truth theories by increasing the number of cycles to 20. 
Overall, these results suggest that \textsc{AutoCog} can recover known ground truth theories and do so even when choices are corrupted by noise. Furthermore, with more cycles, it can overcome its inherent bias towards canonical theories to produce theories that align more with the ground truth. 

\subsection*{Closed loop discovery of theories in multi-attribute decision-making}
Having established that \textsc{AutoCog} can recover known ground-truth theories, we next moved to a closed-loop setting in which \textsc{AutoCog} directly conducts online behavioral studies with human participants to surface theories of decision making (see SI for details).
\textsc{AutoCog} does so by proposing a new experiment every cycle, with a completely new set of option-pairs and validities, and a different set of participants performing each of these experiments. Then, new theories are surfaced by integrating information on how human participants solve different experiment designs across experiments.
As a result, this setting presents a stiff challenge, as it requires surfacing theories that account for multiple sources of variance, namely participant-level differences within each experiment, population-sample differences across experiments, and experimental-design-based differences. 
To this end, we ran \textsc{AutoCog} closed loop for five cycles on a multi-attribute decision-making task, where options take binary-valued expert ratings (Figure \ref{fig:overview}C). Notably, \textsc{AutoCog} conducted ten online behavioral studies and explored five different theories in this setting. 
Each cycle involved 25 participants; with two experiments per cycle, it results in 250 participants in total over five cycles; see Human experiments in the Methods section. For this run, we seeded the system with TTB and WADD with the goal of surfacing a theory that is better than the previous best theory, WADD, in this experimental design space \cite{binzFoundationModelPredict2025, hilbig2014generalized, rmusGeneratingComputationalCognitive2025}.

\begin{figure}
    \centering
    \includegraphics[width=1.04\linewidth]{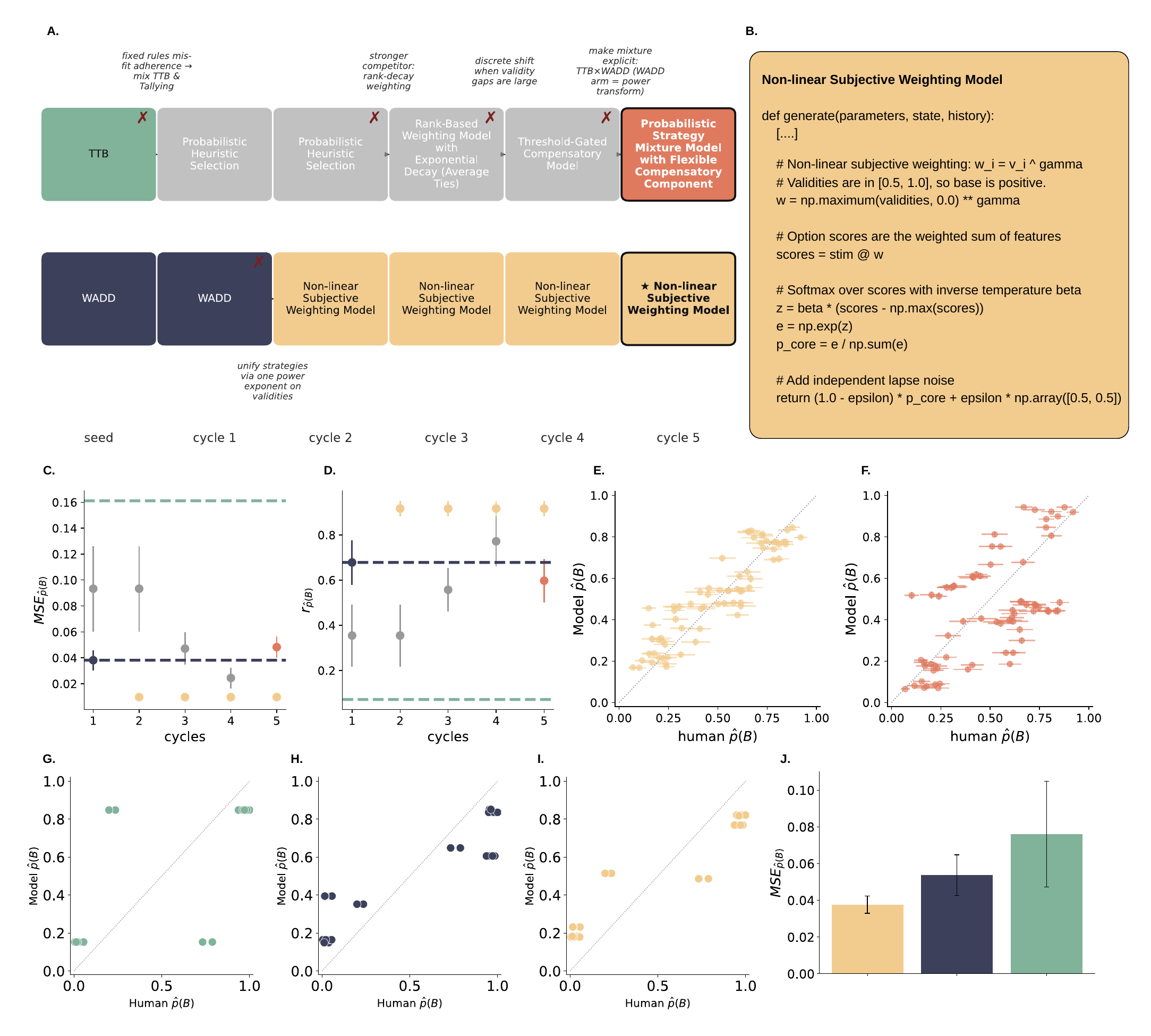}
 
    \caption{\textbf{Closed loop discovery of theories in multi-attribute decision-making tasks} \textbf{(A)}~Theory lineage across the seed and five revision cycles for two theory slots. The top slot churns out different theories while the bottom slot stabilizes on the \emph{Non-linear Subjective Weighting Model} as early as the second cycle and persists with it in every subsequent cycle (starred, gold).  One-line annotations give the arbiter's reasoning at each kill$\rightarrow$admit transition. \textbf{(B)}~\texttt{generate} implementation for \emph{Non-linear Subjective Weighting Model}. \textbf{(C)}~ $MSE_{\hat{p}(B)}$ between model and the human choice data, pooled over all experiments in the run, over cycles (lower $=$ better) and \textbf{(D)}~Pearson correlation $r_{\hat{p}(B)}$ between model and the human choice data, pooled over all experiments in the run, over cycles (higher $=$ better). Models colored as in (A). \textbf{(E, F)}~Human vs.\ model $\hat{p}(B)$ across stimuli for the two final  theories---\emph{Non-linear Subjective Weighting Model} (E, gold) and \emph{Probabilistic Strategy Mixture Model}  (F, peach); points on the diagonal indicate accurate prediction.  \textbf{(G--I)}~Generalization of the surfaced model to unseen human dataset \cite{hilbig2014generalized} shown as human-vs-model $\hat{p}(B)$ scatter for the seeds and the winning surfaced theory, colored as in (A).   \textbf{(J)}~ $MSE_{\hat{p}(B)}$ between model and human responses from the Hilbig et al. study \cite{hilbig2014generalized} for the seeds and winning surfaced theory; black lines indicate mean $\pm$ SEM}
    \label{fig:human1}
\end{figure}

To evaluate the performance of \textsc{AutoCog}-surfaced theories at the population level, we calculated the MSE between the predictions of the model and human choices across all unique pairs of options, pooling data from all experiments conducted in the run ($MSE_{\hat{p}(B)}$). 
%
We found that \textsc{AutoCog}-surfaced theories improve significantly over cycles (Figure \ref{fig:human1}C-D), with $MSE_{\hat{p}(B)}$ for the best surfaced theory decreasing by an order of magnitude, from $0.093$ $(SEM=0.033)$ at the first cycle to $0.0097$ ($SEM=0.0018$) at the end of five cycles. The improvement over cycles is further corroborated by a linear mixed-effects model (random intercept per experiment) that showed a significant effect of the cycle on $MSE_{\hat{p}(B)}$ of surfaced theories ($\beta=-0.014$, $z=-3.24$, $p=0.001$).

A key advantage of \textsc{AutoCog} is that the full trajectory taken to surface these two theories is available for study; see Figure \ref{fig:human1}A for a visual overview of the theories explored in this run. 
At the end of its five cycles, the two theories surfaced by \textsc{AutoCog} are as follows, reproduced verbatim (see SI for full theories): 
\begin{quote}
\textbf{(1)} ``\textit{Non-linear Subjective Weighting Model}: Subjects evaluate
options by computing a weighted sum of their features, but they do not use the
objective cue validities directly. Instead, subjective cue weights are a power
function of the provided validities. An individual-specific exponent parameter
controls the non-linearity of this transformation. This single mechanism unifies
multiple decision strategies: an exponent near 0 flattens the weights (yielding
Equal-Weight/Tallying), an exponent of 1 uses the validities linearly (yielding
WADD), and a large exponent strongly amplifies the most valid cues (yielding
non-compensatory Take The Best behavior)''

\medskip

\textbf{(2)} ``\textit{Probabilistic Strategy Mixture Model with Flexible
Compensatory Component}: Subjects maintain a repertoire of distinct cognitive
strategies---a non-compensatory heuristic (Take-The-Best) and a compensatory
strategy (Weighted Additive). In any given trial, they probabilistically select
which strategy to deploy based on an individual trait parameter. To accurately
capture heavily compensatory behavior observed in certain environments, the
Weighted Additive strategy uses a non-linear power transformation of the
validities, allowing subjects to tune their compensatory integration rather than
strictly using raw validities.''
\end{quote}
Both surfaced theories share a common feature that is missing in the seed theories: a non-linear transformation of the validities. This feature comes with the added benefit of being parsimonious in terms of its number of parameters. For example, the non-linear subjective weighting model uses as little as three parameters. 
In fact, Rmus et al. \cite{rmusGeneratingComputationalCognitive2025}, using an LLM-based program synthesis called GeCCo, and Binz et al. \cite{binzFoundationModelPredict2025}, using human-in-the-loop scientific regret minimization, independently recovered a cognitive model instance analogous to the non-linear subjective weighting model surfaced by \textsc{AutoCog}.\footnote{The cognitive models produced by \cite{rmusGeneratingComputationalCognitive2025, binz2025asmr, binzFoundationModelPredict2025} were made publicly available well after the training data cut-off date (January 2025) of \textsc{Gemini-3.1-Pro} used in \textsc{AutoCog}, so we do not expect there to be any data contamination.}
However, unlike prior approaches, \textsc{AutoCog} surfaced these theories even though (1) there was no parameter optimization to fit to participants' choices; (2) its goal was to capture human behavior across multiple populations and tasks and not explain a single population and a specific task instance; (3) it did not include an explicit pressure for parsimony.

Among the two surfaced theories, the non-linear subjective weighting model ($M=0.0097, SEM=0.0018$) performed significantly ($t(9)=-4.20$, $p=0.002$, $d_z=1.33$) better than the probabilistic strategy mixture model ($M=0.0482, SEM=0.0082$) over ten experiments in terms of $MSE_{\hat{p}(B)}$.
In Figure \ref{fig:human1}E-F, we visualize the scatter plot for $\hat{p}_B$ between surfaced theory and human pooled across experiments and offer additional confirmation that the strong performance of the non-linear subjective weighting model is not confounded by any outliers. 
For an exact implementation of the non-linear subjective weighting model, see Figure \ref{fig:human1}B.
We then tested how well this model generalizes to an unseen decision-making task with binary-valued cues. Specifically, we considered behavioral data from a study more than a decade ago,  Hilbig et al. \cite{hilbig2014generalized} consisting of 24 unique pairs of options with four cues each and fixed validities of 0.9, 0.8, 0.7 and 0.6, respectively.  
As baselines, we considered all models considered in the original study, namely, TTB, Tallying, and WADD. As shown in Figure \ref{fig:human1}J, we found that the non-linear subjective weighting model ($M=0.0377, SEM=0.0047$) performed better than the baselines, TTB ($M=0.0760, SEM=0.0288$), Tallying ($M=0.0784, SEM=0.0201$), and WADD ($M=0.0537, SEM=0.0111$) in terms of $MSE_{\hat{p}(B)}$. 
Likewise, the non-linear subjective weighting model ($M=0.96,SEM=0.02$) performed better than the seed theories (TTB: $M=0.78,SEM=0.11$; WADD: $M=0.91,SEM=0.02$) in terms of $r_{\hat{p}(B)}$ as shown in Figure \ref{fig:human1}G-I.
Together, these results indicate that \textsc{AutoCog} can surface a theory that generalizes to unseen population-sample and task-design beyond those observed within its own cycles.

\subsection*{Extending \textsc{AutoCog} to more complex decision-making tasks}

Next, we expanded the experimental design space so that \textsc{AutoCog} can make contact with a richer space of theories while still maintaining the current task structure and framing. Concretely, we extended the experimental design space from binary expert ratings to cardinal ratings, allowing ratings to take integer values from $0$ to $r_{\max}$ (see Figure~\ref{fig:overview}D). 
Even such a small change in the design space has an effect on the kinds of theory with which the system can connect. For example, heuristic strategies, such as TTB and Tallying \citep{gigerenzer1996reasoning, gigerenzer1999simple, gigerenzer2011heuristic}, are not sensitive to the magnitude of the cue values, whereas models from the multi-attribute utility theory can capture the differences in attribute magnitudes by computing subjective utilities over attribute values \cite{keeney1976decision}.
However, unlike tasks studied under multi-attribute utility theory, our expanded experimental design space assumes validities for each cue are provided to the participant and are not subjective (participant specific) weights. 
Next, we ran \textsc{AutoCog} for five cycles with this expanded experimental design space, while using TTB and Tallying as seed theories and keeping all other experiment parameters matched to the previous run. 

As earlier, we verified whether \textsc{AutoCog} surfaced theories are population-invariant and task-invariant over cycles with a linear mixed-effects model (random intercept per experiment). It showed a significant effect of the cycle on $MSE_{\hat{p}(B)}$ of surfaced theories ($\beta=-0.0063$, $z=-4.45$, $p<0.0001$); see also Figure \ref{fig:human2}C-D.
In terms of the discovery trajectory, \textsc{AutoCog} discarded both seeds within the first two cycles, then fully explored within the WADD family after that (see Figure \ref{fig:human2}A for a detailed visualization).
The following are two theories surfaced by \textsc{AutoCog} at the end of five cycles, reproduced verbatim (see SI for full theories): \begin{quote}
\textbf{(1)} ``\textit{Diminishing Returns WADD:} This theory posits that individuals evaluate options by applying a concave utility function to cardinal cue values before weighting them by their cue validities. By compressing large cardinal values via a power law transformation, extreme advantages on a single cue are discounted relative to multiple moderate advantages across several cues. Parameterizing the shift applied before the power law allows the model to flexibly smooth the extreme marginal utility near zero, preventing over-sensitivity to small integer differences while maintaining the core concave utility mechanism.''

\medskip

\textbf{(2)} ``\textit{Threshold-based Binarization (Satisficing WADD):} Decision-makers simplify complex cardinal information by converting continuous or multi-level ratings into binary cues based on a satisficing threshold. A feature is considered satisfactory if its rating meets or exceeds the threshold and unsatisfactory otherwise. The options are then evaluated by computing the validity-weighted sum of these binarized features (WADD on binary cues). This mechanism naturally explains why extreme cardinal advantages (e.g., 10 vs 5) might be ignored if both options exceed the satisficing threshold, allowing an option with distributed moderate advantages to win against an option with a single extreme advantage.''
\end{quote}

As shown in Figure \ref{fig:human2}C, the \textsc{AutoCog}-surfaced theories ($M=0.018$, $SEM=0.004$; paired $t(9)=-11.8$, $p<10^{-6}$, $d_z=3.73$) were consistently better than the seed theories ($M=0.138$, $SEM=0.009$) across the ten experiments collected in the run, in terms of $MSE_{\hat{p}(B)}$.
The pooled performance of the seed theories in the run was as follows: Tallying ($M=0.1480, SEM=0.0237$) and TTB ($M=0.1283, SEM=0.0266$).
Among the two surfaced theories, the Diminishing Returns WADD ($M=0.0180,SEM=0.0051$; $t(9)=-0.20$, $p=0.85$, $d_z=0.06$) was only marginally better than Satisficing WADD ($M=0.0180,SEM=0.0033$). 
While in terms of $r_{\hat{p}(B)}$, the two surfaced theories were also indistinguishable (Satisficing WADD: $M=0.80$, $SEM=0.04$; Diminishing-Returns WADD: $M=0.77$, $SEM=0.08$; paired $t(9)=0.56$, $p=0.59$, $d_z=0.18$).
However, when we visualized the scatter plots for $\hat{p}(B)$ between the surfaced theory and humans as previously (Figure \ref{fig:human2}E-F), it became apparent that the high $r_{\hat{p}(B)}$ observed in the case of the Satisficing WADD ($72/134=54\%$) is confounded by significant predictions $\hat{p}(B)\in[0.45,0.55]$ compared to Diminishing Returns WADD ($27/134=20\%$). 
As a result, we chose Diminishing Returns WADD to be the winning theory surfaced by \textsc{AutoCog} for this run; see Figure \ref{fig:human2}B for its exact model implementation.

\subsubsection*{Testing the proposed theory}

Next, we designed a follow-up study where we pre-registered the predictions made by Diminishing Returns WADD (the winning surfaced theory from \textsc{AutoCog} for this experimental setting) including the analysis scripts, and compared them against human responses (\href{osf-preregistration}{https://osf.io/f7kes}). The study tested the following three hypotheses (see Figure \ref{fig:human2}G-J):

\begin{itemize}
\item \textbf{H1: Model-discrimination hypothesis.} On stimuli constructed to separate Diminishing Returns WADD from a given alternative model, participants should choose the option predicted by Diminishing Returns WADD more often than the option predicted by the alternative model.

\begin{figure}
    \centering
    \includegraphics[width=1.0\linewidth]{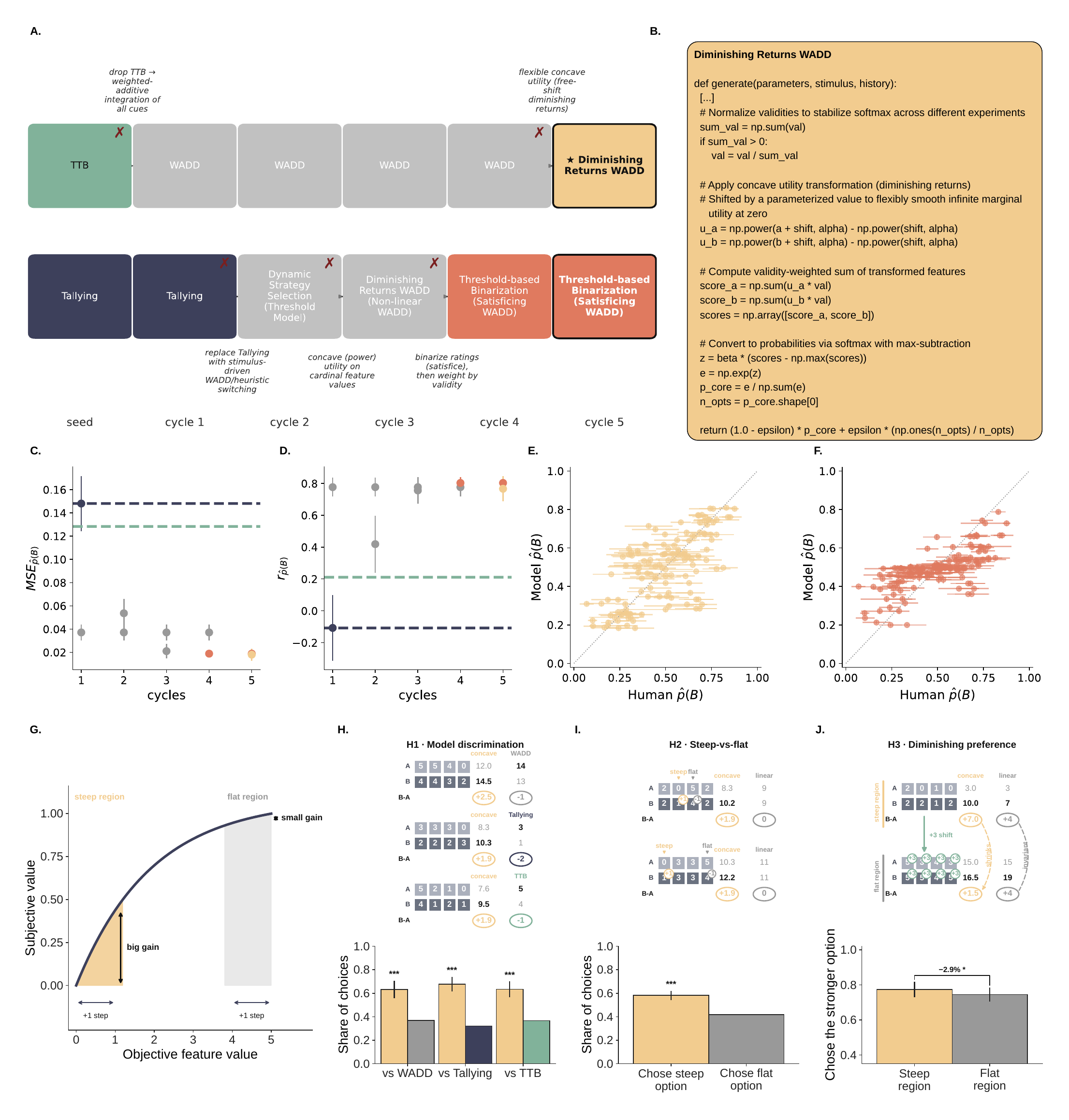}

  \caption{\textbf{End-to-end discovery of a
  Diminishing Returns WADD model.}
  \textbf{(A)}~Theory lineage over seed plus five cycles. The top slot evolves
  TTB $\rightarrow$ WADD $\rightarrow \cdots \rightarrow$ \emph{Diminishing Returns
  WADD} (starred), in which cue values pass through a concave utility transform;
  the bottom slot evolves Tallying $\rightarrow \cdots \rightarrow$
  \emph{Threshold-based Binarization (Satisficing WADD)}, which binarizes
  ratings by validity. Transition annotations give the arbiter's reasoning.
  \textbf{(B)}~The \texttt{generate} implementation of the Diminishing Returns
  WADD model: validities normalized for cross-experiment comparability, a
  parameterized concave transform of cue values giving diminishing marginal
  utility, a validity-weighted sum, and softmax choice with lapse noise.
\textbf{(C)}~$MSE_{\hat p(B)}$  between model and the human choice data, pooled over all experiments in the run, over cycles (lower $=$ better) and \textbf{(D)}~Pearson correlation $r_{\hat p(B)}$ between model and the human data, pooled over all experiments in the run, over cycles (higher $=$ better)
  \textbf{(E, F)}~Human vs.\ model $\hat{p}(B)$ scatter for the two surviving
  theories.
  \textbf{G-J}~Mechanism and three preregistered behavioral tests of the
  discovered concave subjective-value account: the inferred utility curve (large
  gains in the low-value region, small gains in the high-value region), and
  tasks H1 (model discrimination---does the model out-predict WADD, Tallying,
  and TTB?), H2 (steep- vs.\ flat-region choices), and H3 (diminishing
  preference between the stronger option in a steep vs.\ flat region), each with
  example stimuli and the corresponding share/choice bar charts (error bars
  SEM; significance markers shown).}
    \label{fig:human2}
\end{figure}

\item \textbf{H2: Steep-vs-flat hypothesis.} Participants should choose an option with an advantage in the low-value region more often than an option with an objectively equal-sized advantage in the high-value region.

\item \textbf{H3: Level-shift hypothesis.} The same cue-value structure should have a stronger effect when presented in a low value range than when shifted upward into a higher value range.

\end{itemize}

We tested H1 using model-discriminating stimuli. For each alternative model, stimuli were constructed so that Diminishing Returns WADD and the alternative model made opposing predictions. For each participant and each comparison, we computed the proportion of choices matching Diminishing Returns WADD and the proportion matching the alternative model. The preregistered analysis used one-sided paired $t$-tests to test whether the Diminishing Returns WADD match rate exceeded the alternative-model match rate. This was done separately for WADD, tallying, and take-the-best, with the three planned comparisons corrected using the Holm procedure.

We tested H2 using steep-vs-flat trials. These trials contrasted equal-sized cue value steps that occurred in different regions of the value scale. Under a concave value function, the lower part of the scale is the \textit{steep} region, where a small increase in objective cue value should produce a relatively large increase in subjective value. The upper part of the scale is the \textit{flat} region, where the same objective increase should produce a smaller increase in subjective value. For example, one trial could compare option A = $[1, 4, 2, 2]$ with option B = $[0, 5, 2, 2]$. Option A has a one-unit advantage in the low range on the first cue, $1$ vs. $0$, whereas option B has an equal one-unit advantage in the high range on the second cue, $5$ vs. $4$. A linear model treats the low-range and high-range steps as equivalent, while Diminishing Returns WADD predicts that the low-range step should have a greater impact because it lies on the steeper part of the concave value function. For each participant, we computed the proportion of steep-vs-flat trials on which they chose the low-range advantage option. The preregistered analysis used a one-sided one-sample $t$-test against $0.5$ to test whether this proportion was above chance.

We tested H3 using level-shift trials. These trials asked whether the same objective cue-value structure has a weaker effect when shifted upward into a higher part of the value scale. Under a linear model, adding a constant to the relevant feature values should not change the effect of the structure, because the objective differences between the options remain the same. Under Diminishing Returns WADD, however, the shifted-up version should have a weaker effect because the same objective differences now fall on a flatter part of the concave value function. For example, a low-range trial could compare option A = $[2, 0, 1, 1]$ with option B = $[0, 2, 1, 1]$, while the shifted-up version of the same trial could compare option A = $[5, 3, 4, 4]$ with option B = $[3, 5, 4, 4]$. The objective structure is identical in both cases: option A has a two-unit advantage on the first cue, and option B has a two-unit advantage on the second cue. The only difference is that the entire structure has been shifted upward. A linear WADD model treats the low-range and shifted-up versions as equivalent, whereas the Diminishing Returns WADD predicts that the contrast should be stronger in the low-range version because the same two-unit steps occur on the steeper part of the value function. For each participant, we computed the difference between the proportion of target-option choices in the low-range version and the proportion of target-option choices in the shifted-up version. The preregistered analysis used a one-sided one-sample $t$-test against $0$ to test whether this offset effect was positive.

The preregistered results followed the predicted pattern (Table~\ref{tab:preregistered-followup}). For H1, participants matched Diminishing Returns WADD more often than each alternative model on the corresponding model-discriminating trials. This held for WADD, $M_{\Delta} = 0.264$, $SE = 0.076$, $t(49) = 3.465$, one-tailed $p = 5.56 \times 10^{-4}$, $d = 0.490$; tallying, $M_{\Delta} = 0.356$, $SE = 0.063$, $t(49) = 5.665$, one-tailed $p = 3.83 \times 10^{-7}$, $d = 0.801$; and take-the-best, $M_{\Delta} = 0.268$, $SE = 0.068$, $t(49) = 3.927$, one-tailed $p = 1.35 \times 10^{-4}$, $d = 0.555$. All three comparisons survived Holm correction.
For H2, participants chose the low-range advantage option above chance in the steep-vs-flat trials, $M = 0.582$, $SE = 0.020$, 95\% CI $[0.543, 0.621]$, $t(99) = 4.192$, one-tailed $p = 3.01 \times 10^{-5}$, $d = 0.419$. 
For H3, participants showed the predicted level-shift effect: target-option choices were higher in the lower-range version ($M = 0.774$) than in the shifted-up version ($M = 0.745$), yielding a positive offset effect, $M_{\Delta} = 0.029$, $SE = 0.016$, 95\% CI $[-0.003, 0.061]$, $t(99) = 1.822$, one-tailed $p = .036$, $d = 0.182$.

Together, these results show that Diminishing Returns WADD generated prospective predictions that held in a preregistered follow-up study. This provides a preregistered test of the theory surfaced by \textsc{AutoCog}, rather than an additional post-hoc comparison within the original discovery cycle.

\section*{Discussion}
Across the empirical sciences, the discovery cycle is increasingly carried out by autonomous systems that design experiments, collect data, compare competing explanations, and revise their theories \citep{musslickAutomatingPracticeScience2025, novikov2025alphaevolve}. Psychology is well placed to do the same: experimental design, online data collection, and model synthesis have all matured, and powerful AI systems can now connect them \citep{hartshorneCriticalPeriodSecond2018a, peterson2021using, cavagnaroAdaptiveDesignOptimization2010}.
The last step needed to close the loop is the generative work of theory-building, which has remained a largely manual, creative act \citep{eronenTheoryCrisisPsychology2021, muthukrishnaProblemTheory2019}. Here, we present \textsc{AutoCog}, a fully autonomous agentic system that closes the loop on human behavior by automating exactly this step.
A researcher provides four inputs: an experiment design space, a model design space, the two seed theories that start the search, and a behavior source; the loop then searches over theories and experiments, designing discriminating studies, gathering behavioral data, diagnosing why the weaker account fails, and synthesizing its successor. To our knowledge, this demonstration is the first of its kind in two respects: (i) every stage between proposing an experiment and revising a theory, including the collection of human data, runs using an LLM agent without researcher intervention; and (ii) scientific discovery operates at the level of theories, natural-language mechanistic claims paired with executable models, while other automated discovery methods have only searched spaces of model forms over fixed datasets \citep{petersonUsingLargescaleExperiments2021a, rmusGeneratingComputationalCognitive2025, castroDiscoveringSymbolicCognitive2025}. Our results demonstrate that this first instance already performs well in the classical domain of human decision-making.

Several of our results establish that the loop searches theory space effectively. When behavioral data were generated from known models, \textsc{AutoCog} recovered the ground-truth theory across noise levels despite being seeded with theories from different families (Figure~\ref{fig:recovery}). This recovery extended to hand-crafted mechanisms that were constructed to be unconventional including strategies that choose against the majority of decision-making strategies or perseverate on a single option: most were recovered within five cycles, and the more elusive ones were within reach when the loop was given additional cycles. 
In addition, when given pure noise it converged on theories equivalent to random guessing rather than imposing structure on it.  These controls indicate that what \textsc{AutoCog} surfaces is driven by the collected behavioral data rather than by what its components find easy to express (see SI for examples).

Furthermore, the surfaced theories improved over cycles and predicted behavior beyond the data that produced them (see Figure \ref{fig:illustrativecasestudy} for a demonstration). In the human runs, surfaced theories captured behavior better than either seed on all ten collected datasets, improving over early cycles and stabilizing thereafter, and generalized to a held-out experiment from the literature, where they again outperformed the seed theories.
The theory surfaced by \textsc{AutoCog} changed with the structure of the task. In the binary-feature task, the winning theory unified the three canonical heuristics, recovering Tallying, WADD, and Take-the-Best as limiting cases of a single subjective-weighting mechanism governed by one free exponent (Figure~\ref{fig:human1}). In the cardinal-feature task, the loop surfaced a regularity we did not anticipate: people pass cue magnitudes through a concave, diminishing-returns valuation before weighting by reliability, so a given advantage counts for more when the values are small than when they are large. This reflects how the \textsc{AutoCog} loop carries out open-ended discovery by generating novel theories rather than merely choosing among a predefined set. None of the canonical heuristics predicts this curvature, so it was a genuinely new prediction. We tested it prospectively in a preregistered study on new participants, and it held: choice showed diminishing sensitivity to feature value where the heuristics predict none (Figure~\ref{fig:human2}). Independent of the system that surfaced it, the regularity is a new and confirmed fact about how people decide, and one that echoes the diminishing sensitivity that is central to prospect theory.

These results indicate what theorizing can look like when the generative step is automated. Every stage of a run, the proposed experiments and metrics, the observed data, the arbiter's verdicts, and the synthesized theories, is logged as a machine-readable trace, so a discovery run becomes an auditable research artifact \citep{liu2026last}. The path taken by the system is therefore open to inspection: another researcher, or another agent, can see why each candidate was set aside and resume the search from its frontier rather than restarting it if they so desire. 
Unlike other automated discovery systems, humans play a key role in \textsc{AutoCog}.
Their contribution shifts from execution to specification, from running studies to deciding what counts as a good theory, what properties it should satisfy, what form it may take, and what space to search. 
Running the loop against a behavioral foundation model \cite{binzFoundationModelPredict2025} rather than people points to a further practice within reach, in which initial searches are run in-silico and human participants are reserved for the candidates that survive \cite{jagadishCanWeAutomatize2026}, though how far this extends depends on how faithfully the foundation model captures human behavior in the target domain. 

The outcomes of \textsc{AutoCog} are still bounded by constraints imposed by the human researcher. By design, it searches only within the experiment design space and the model design space the agents are tasked to explore, and a theory advances by matching observed behavior across the accumulated data: parsimony, unification, and interpretability, the other virtues of a good theory \citep{guestMartinForceTheory2021, vanRooijBaggioTheoryBeforeTest2021}, are not scored as explicit objectives. Even so, because a candidate progresses only by capturing all the behavior collected so far, across populations and tasks, the demand to generalize acts as an implicit pressure toward these virtues, as the unifying account that won the binary-feature task suggests. Other limitations are inherited: The loop verifies that each synthesized model compiles and predicts the data, but not that its code faithfully realizes the verbal theory it is paired with, a verbal-to-formal translation problem the field is yet to solve. Its language models are biased toward mechanisms common in their training data but a candidate is retained only after it survives experiments that simulation confirms can separate the competing theories, and captures observed behavior better than they do. This enables \textsc{AutoCog} to recover mechanisms its models would not otherwise readily volunteer. The novelty we observed was not a deliberate target: the loop rewards capturing behavior, not exploring diverse or unfamiliar accounts, so novelty emerged rather than being sought. A future system could pursue it directly by optimizing the diversity score of the proposed theories \citep{friedman2022vendi}
Adversarial testing and revision is just one of several procedures for surfacing theories, and pitting competing accounts against discriminating experiments has a long history both in scientific method broadly \citep{platt1964strong, godara2026adversarial} and in the testing of cognitive theories \citep{meehl1978theoretical, robertsPashler2000}. Other systems might forgo this adversarial tension, for instance by triangulating converging evidence from independent methods  \citep{munafo2018robust}, by unifying phenomena under a shared  mechanism \citep{newell1990unified}, or by progressively refining a single account \citep{lakatos1970}. Each could be pursued within \textsc{AutoCog}. Whichever of these approaches  we adopt, the larger point remains the same: discovering psychological theories of human behavior is now within the reach of automation.


\section*{Methods}

\subsection*{Task Domain}

We studied human decision-making with a two-alternative multi-attribute decision-making task adapted from the protocol used in \cite{hilbig2014generalized}. Here, we describe the task used for the expanded experimental design space with cardinal cue values between $0$ and $r_{\max}$. The corresponding task with binary cue values was identical in structure. 

In each trial, the participants chose which of the two fictitious products, A and B, was of higher quality. Each product was described by a vector of expert ratings (cue values). Ratings were integers on a $0$--$r_{\max}$ scale, displayed as horizontal filled bars annotated with their numeric value (e.g.\ ``4/5''). Each expert was associated with a \emph{validity} that was fixed throughout the experiment and described to the participant as the probability that the expert's rating points to the true quality of the product. Validities were communicated to participants up front in an ``Expert accuracies'' panel and were also displayed as a percentage beside each rating on every trial (e.g.\ ``Expert~1 (95\%)''). The number of experts, their validities, the rating ceiling $r_{\max}$, and the set of product ratings (option-pair) were not fixed by the experimenters: they were proposed automatically by \textsc{AutoCog} following the principle of adversarial experimental design (Stage~1, experimental design).

On each trial, participants were prompted to indicate their choice by pressing ``A'' or ``B''. There was no time limit and no trial-by-trial correctness feedback. To ensure that participants inspected the full stimulus before responding, the prompt to respond was hidden, and the response keys were locked for the first $1500$\,ms of each trial; the prompt to respond with ``A''\,/\,``B'' appeared only afterwards. Each participant completed approximately $96$ trials in a single block \citep{hilbig2014generalized}: every unique product pair in the design was repeated $K = \max(1, \lfloor 96 / n_{\text{pairs}} \rfloor)$ times, and the trial order was randomized independently for each participant. At the end of the task, participants saw a summary screen reporting their accuracy relative to a Bayesian normative benchmark, computed as the log-odds choice rule in which each expert contributes $(a_i - b_i)\log\!\big(v_i / (1 - v_i)\big)$ where $a_i$ and  $b_i$ are the two products' ratings on expert $i$ and $v_i$ is that expert's validity.


\subsection*{AutoCog}

\textsc{AutoCog} is an automated pipeline for adversarial scientific discovery in which two LLM agents iteratively propose, test, and revise theories of behavior. 
It was designed to (1) be natively agentic, (i.e. set up to improve as agentic harnesses get better); (2) require minimal but essential human intervention (i.e. researchers specify factors shown in Figure \ref{fig:overview}D once and it runs autonomously after): (3) use iterative self-verification (i.e. in each stage LLM proposals, such as theories, models, experiments, and metrics are iteratively refined through self-verification); (4) be fast and reliable (i.e., simulation based inference for model comparison instead of model fitting, avoiding problems of overfitting and long fitting times); (5) interpretable (e.g., reasoning trajectory for proposing an experiment or a model are logged).

In this study, we used the same pipeline across two sources of ground-truth behavior in multi-attribute decision-making tasks: specified, known cognitive theories and human participants recruited online. After initialization (0), each cycle consists of adversarial experimental design (1), data collection (2), theory analysis and arbitration (3), and theory revision (4). The pipeline operates on executable research artifacts: theories can be simulated to generate predictions, experiment specifications can be run on theories or sources of behavior, analysis procedures can evaluate simulated and observed data, and regenerated theories can be tested again in subsequent cycles. These artifacts allow each cycle to connect experimental design, behavioral observation, model comparison, and theory revision within a single closed loop. In the following, we describe one cycle. Cycles are chained so that later theories are evaluated against the evidence accumulated in earlier cycles (see SI for visualization of each cycle).

\subsubsection*{Pipeline}

\begin{itemize}

\item[0.]  \textbf{Initialization:} Each run begins with two competing theory slots. Each slot contains an executable theory: a natural-language description, a predict function mapping stimuli and trial history to response probabilities, a policy function mapping those probabilities to choices, and free parameters with admissible ranges. The initial theories depended on the source of ground-truth behavior. In known-theory runs, the seed theories were chosen to exclude the true data-generating theories. In human-participant runs, the two seed theories were TTB and Tallying for the run with cardinal cue values, and TTB and WADD for the run with binary cue values. 

\item[1.]  \textbf{Experimental design:} In each cycle, both agents act as experiment proposers. Each agent proposes an experiment under which its own theory should outperform its competitor, together with a computable metric intended to quantify that advantage. In this study, experiments were multi-attribute decision-making tasks. In each trial, a subject chose between two options, A and B, each described by a vector of integer expert ratings. The specification of the experiment proposed by the agent defined the design-level structure of the task: the number of experts, encoded by the length of the validity vector; the validity assigned to each expert; the rating scale; and a set of unique A/B rating pairs. These quantities were fixed within an experiment. Together with the proposed metric, which maps the resulting choice data to a scalar outcome, this formed an experiment--metric pair.

Because theories are executable, \textsc{AutoCog} validates proposed pairs before data collection using the same forward-simulation logic used for later theory analysis. For each proposed pair, \textsc{AutoCog} samples parameter values from each theory's admissible ranges, simulates the experiment under each sampled parameterization, and applies the proposed metric to the simulated data. A pair is accepted only if the resulting metric values distinguish the two theories at the planned sample size ($N=25$ subjects). \textsc{AutoCog} then applies a Welch two-sample test parameterized by the planned sample size; pairs are accepted when the two-sided $p$ value is below $\alpha=.01$. Each agent attempts up to three experiments and up to four metrics per experiment. The first accepted pair is committed; if no pair is accepted, the final attempted pair is committed. Thus, LLM agents propose theoretically targeted tests, while the executable pipeline verifies that those tests induce discriminable predictions before data collection.

\item[2.]  \textbf{Behavioral data collection:} Each committed experiment is run on the source of ground-truth behavior, producing observed data. In known-theory runs, behavior was generated by a specified cognitive model, allowing us to test whether \textsc{AutoCog} could recover the data-generating process. In human-participant runs, behavior was generated by participants recruited online. For each committed experiment, the unique A/B rating pairs were repeated as needed and randomized to produce the subject-level trial schedule. For human-participant runs, these schedules were rendered as online experiments using SweetBean \cite{sweetbean} and deployed through AutoRA \cite{Musslick_AutoRA_Automated_Research_2024}. Participants were recruited through Prolific ($N=25$ per experiment; approximately $6$--$8$ minutes; compensation US\$0.80).

\item[3.]  \textbf{Analysis \& Arbitration:} After data collection, \textsc{AutoCog} scores theories by forward simulation rather than by fitting them separately to each observation, using the same procedure described above for validating proposed experiment--metric pairs. For each committed experiment, the metric is applied to the observed data to obtain the empirical outcome, and to each theory's simulated data to obtain that theory's predicted outcome. This generative evaluation was used because \textsc{AutoCog} evaluates each theory together with its admissible parameter ranges: the theory should not only capture the behavioral target specified by the proposed metric, but also generate a distribution of simulated participants that can be compared with the observed distribution. Because the metric is evaluated on the observed data, each experiment--metric pair yields an empirical target that is independent of the rationale for which the metric was originally proposed. This allows \textsc{AutoCog} to score every theory on every committed experiment--metric pair, including pairs proposed by competing theories. For arbitration, \textsc{AutoCog} presents the analysis agent with the committed experiment--metric pairs, including the experiment description, the metric description, the observed metric value, and the corresponding predicted metric values generated by each competing theory. The analysis agent uses this information to compare how well the competing theories account for the empirical outcomes across the committed experiments, producing scores that determine which theories are retained, revised, or replaced in the next cycle.

\item[4.]  \textbf{Theory revision:} The arbitration verdict determines how the targeted theory slot is updated. If the verdict calls for a revised model, the natural-language theory description is retained and only the executable components are regenerated. If the verdict calls for a revised theory, both the natural-language description and the executable model are replaced. Candidate revisions are generated in an inner critique loop of up to ten iterations. In each iteration, a candidate is compiled and validated, with up to ten retries if needed, then simulated on the existing pool of committed experiment--metric pairs and evaluated against the empirical outcomes using the same forward-simulation procedure described above. A candidate is retained only if it improves on the best candidate encountered so far, where improvement is defined as a lower aggregate loss: the mean discrepancy between the observed metric values and the corresponding metric values predicted from simulated data. If no candidate is explicitly accepted during the critique loop, the candidate with the lowest aggregate loss is retained. The revised theory is then pinned to its slot, its predictions are recomputed for the existing committed experiment--metric pairs, and the next cycle begins.

\end{itemize}

\subsubsection*{Design space}

\paragraph{Experimental design space.} For each proposed experiment, we held certain task parameters fixed and allowed \textsc{AutoCog} to propose design parameters that can distinguish competing theories along dimensions that the researcher thinks are interesting. Fixed across all proposals were the choice structure (a forced binary choice between two products A and B on every trial), the absence of trial-by-trial feedback, the trial budget (maximum of 100 trials per subject, after \citep{hilbig2014generalized}), and the randomization of trial order independently per subject. Within this protocol, \textsc{AutoCog} proposed four design parameters per experiment: the number of expert cues (set implicitly by the length of the validity vector), the per-cue validities (each constrained to [0.5, 1.0], with the requirement of a non-degenerate spread, since uniform validities render weighted-additive and equal-weight rules behaviorally identical), the rating scale ceiling r (an integer $\geq$1, where r = 1 recovers the classic binary-cue protocol and r $\geq$ 2 yields cardinal cues that dissociate magnitude-based from sign-based heuristics), and the set of stimulus pairs themselves (integer rating vectors in [0, r] for each option). A schema-level validator enforced these bounds and internal consistency before any proposal was simulated, so that the free parameters were exactly the levers controlling theory discriminability while all potential confounding conditions were held constant.

\paragraph{Model design space.} A candidate theory is operationalized as executable code \cite{rmusGeneratingComputationalCognitive2025, xie2026think}. We fixed the interface that a theory must satisfy and let \textsc{AutoCog}  produce the mechanism within it. Fixed by the framework was the runtime contract: every theory exposes a predict(parameters, state, history) function returning a choice-probability vector over the two options and a policy(probs) function mapping those probabilities to a response, and parameters are drawn once per simulated subject from their declared ranges rather than fit to data by maximum likelihood. Within this contract, \textsc{AutoCog}  produced the decision rule itself: the Python implementation of predict, the prose statement of the theoretical claim, and the parameter specification (which free parameters the theory has and over what ranges they are sampled). Each theory was equipped with a common stochastic choice model: a softmax over its theory-specific option scores with inverse-temperature $\beta$ (sampled from [0.1, 20.0], spanning nearly random to nearly deterministic response) mixed with a uniform lapse rate $\epsilon$ (sampled from [0.0, 0.5]), together with any theory-specific parameters such as per-cue validities or weights. This separation allowed \textsc{AutoCog} to vary the cognitive mechanism, while the response-generation process and parameter-sampling regime were kept constant across all candidate theories.

\subsubsection*{Implementation}

All stages of the \textsc{AutoCog} discovery loop---experimental design, model analysis and arbitration, and model revision---were driven by a single LLM, Google \texttt{gemini-3.1-pro-preview}, accessed through the \texttt{google-genai} SDK. We used a sampling temperature of $0.7$, a maximum output length of $32{,}768$ tokens, and an extended reasoning (``thinking'') budget of $8{,}096$ tokens. All model calls returned structured JSON enforced through a Pydantic response schema (Gemini native JSON mode), so that proposed experiments, theories and their executable \texttt{predict}/\texttt{policy} functions could be parsed deterministically. Transient network and API failures were handled by an exponential-backoff retry layer (up to $10$ attempts, $2$\,s base delay capped at $120$\,s, with jitter) to tolerate multi-hour runs on the compute cluster.

The pipeline exposes a provider-agnostic LLM interface; in addition to the Gemini models used for the reported results, it supports Anthropic (e.g.\ \texttt{claude-haiku-4-5}), OpenAI (\texttt{gpt-4o}; reasoning models \texttt{o1}/\texttt{o3}/\texttt{o4}) and Princeton AI Sandbox (Portkey gateway) back ends. Reported experiments used Gemini exclusively, with the model identifier recorded in each run directory name.

\subsection*{Baseline theories}
We evaluated recovery against nine decision-making theories: Three are canonical heuristics spanning the frugality--compensation spectrum: \emph{Take The Best} (one-reason choice on the highest-validity discriminating cue), \emph{Tallying} (choose the option winning on more cues), and \emph{Weighted Additive} (choose the higher validity-weighted sum). Six are non-canonical baselines that probe distinct failure modes: \emph{Take The Worst} (one-reason choice on the lowest-validity discriminating cue), \emph{Single-Cue} (choice on the single least-valid cue, with no fallback), \emph{Anti-Majority} (target the option the TTB/Tallying/WADD majority rejects), and three stimulus-independent sequential baselines: \emph{Perseveration} (repeat the previous choice), \emph{Alternation} (switch from it), and \emph{Random} (picks random choices). Except for the two parameter-free sequential baselines, each theory adds choice noise via a softmax over its score (inverse temperature $\beta$) and a uniform lapse (probability $\epsilon$); see SI for details.

\subsection*{Ground-truth recovery}

\paragraph{Ground truths and data generation.} We assessed whether \textsc{AutoCog} recovers the data-generating strategy for two sets of ground truths in the binary decision making task. The first comprised three \emph{canonical} decision heuristics---take-the-best (TTB), weighted-additive (WADD), and tallying---each instantiated as a probabilistic (``sampling'') variant. For these, synthetic choice data were generated at several action-noise levels via an $\epsilon$-greedy rule: each response was replaced by a uniform $50/50$ coin with probability $\epsilon$. We swept $\epsilon \in \{0, 0.5, 0.75\}$. The second set comprised six \emph{non-canonical} baseline strategies, generated without action noise ($\epsilon = 0$): two contrarian strategies (\emph{anti-majority} and \emph{take-the-worst}), one restricted-cue strategy (\emph{single-cue}), and three stimulus-independent serial strategies (\emph{alternating}, \emph{perseveration}, and \emph{random}).

\paragraph{Recovery in choice-proportion space.} Recovery was quantified on the fixed  stimulus set from \cite{hilbig2014generalized}: the 24 unique option pairs defined over four binary cues with validities $(0.9, 0.8, 0.7, 0.6)$. For each theory, we replayed it on these stimuli and computed, per stimulus pair, the
proportion of choosing option~B across $N = 500$ simulated subjects, with each subject's parameters drawn from the theory's prior and each per-stimulus response a probability-matching Bernoulli draw from the theory's predicted $P(B)$. This replay was noise-free: the action-noise level $\epsilon$ indexes the data-generating process that produced the theories, not the recovery scoring itself. History-dependent strategies (alternating, perseveration) were replayed through the full trial sequence, so that the response history they condition on was available; all other strategies used independent per-stimulus draws. All draws were seeded for reproducibility, with independent seed streams for the compared theories and the reference.

\paragraph{Compared groups and reference.} We compared three sets of theories:
the \emph{seed} theories \textsc{AutoCog} was initialized with (round-0 competitors), the
\emph{surfaced} theories it discovered (candidates surviving the final round
plus the final replacement), and the single \emph{best surfaced} theory per run
(selected by the reported metric and averaged across runs). Each was compared to
the \emph{clean ground truth}---the data-generating strategy replayed under the
same noise-free procedure---using mean squared error (MSE) and Pearson
correlation between per-stimulus choice-proportion vectors.

\subsection*{LLM-as-judge}
\textbf{Quantifying theory recovery via an LLM judge.} To assess how well the discovered theory-descriptions recovered the data-generating (``ground-truth'') heuristic, we scored the mechanism similarity between each surfaced theory and the ground-truth theory using an LLM judge (\textsc{Gemini gemini-3-flash-preview}). The judge was prompted (see SI for exact prompts) to act as a cognitive scientist and to rate, on a continuous scale in [0, 1], how similar the two models' underlying decision mechanism is — the rule used to map inputs to a choice — explicitly disregarding wording, terminology, framing, and parameter values. The scale was anchored: 1.0 = the same decision rule (identical choices for the same reason), 0.5 = a recognizable variant or partial overlap (shared core structure with non-trivial deviation), and 0.0 = an unrelated mechanism. The judge returned a JSON object with the score and a one-sentence rationale.

\textbf{Input modes.} We applied the judge to a combination of description and code, where both the written explanation of the theory and the predict() implementation are used to distinguish the mechanisms that are stated from those that are actually implemented. To reduce sampling variance, each theory was scored three times and the scores were averaged.

\textbf{Design and data.} We applied this analysis to the three canonical heuristic families — take-the-best (TTB), tallying, and weighted-additive (WADD) — each used as a ground truth at three action-noise levels ($\epsilon \in  \{0.0, 0.5, 0.75\}$). For every (family, noise) cell we ran the discovery procedure (theories generated by gemini-3.1-pro-preview) five times, each run yielding two surfaced theories, giving 10 surfaced theories per cell and 30 per noise level pooled across families.

\textbf{Aggregation.} For each noise level and input mode we report the mean similarity pooled across all surfaced theories from the three families (each theory weighted equally; $N = 30$), with error bars showing the standard error of the mean. 

\subsection*{Closed-loop theory building with humans}

We applied the same closed-loop discovery procedure to human-participant data. Unlike the ground-truth recovery analyses, the human runs did not assume a known data-generating strategy; instead, \textsc{AutoCog} iteratively proposed experiments, collected human choice data, scored competing theories against the observed outcomes, and revised theories across cycles. For the binary decision-making task, the system was initialized with Tallying and Take-the-Best as the two seed theories. For the extended decision-making task with cardinal feature values, the system was initialized with Take-the-Best and Weighted Additive as the two seed theories.

\paragraph{Evaluation of surfaced human theories.} We evaluated surfaced theories using held-out diagnostic tasks matched to the corresponding design space. For theories surfaced in the binary decision-making task, we used the same fixed  stimulus set from \cite{hilbig2014generalized} as in the ground-truth recovery analyses: 24 unique option pairs defined over four binary cues with validities $(0.9, 0.8, 0.7, 0.6)$. For theories surfaced in the extended design space with cardinal feature values, we evaluated predictions against a preregistered study designed to test diagnostic consequences of diminishing returns in multi-attribute choice. This study tested whether participants preferred options predicted by the Diminishing Returns WADD model over alternatives on model-discriminating stimuli, whether advantages in low-value regions had stronger effects than objectively equal advantages in high-value regions, and whether the same feature-value structure exerted a stronger influence when embedded in a lower rather than higher value range.

\paragraph{Choice-proportion fit.} For each surfaced theory and each of the ten
experiments in a run, fit was measured in choice-proportion space. We summarized
each unique A/B rating pair by $\hat p(B)$, the proportion of B choices: for
human data, pooled over the $N\!=\!25$ participants of that experiment; for a
theory, by replaying it on the same pairs across $N=200$ simulated participants
(each drawing parameters from the theory's prior and emitting a
probability-matching Bernoulli response from its predicted $P(B)$) and taking the
per-pair mean. All draws were seeded (\texttt{base\_seed}$=0$) with a shared seed
stream across theories, so differences reflect mechanism rather than sampling.
Per experiment we computed $MSE_{\hat p(B)}$ between the model and human per-pair vectors. ``Pooled
over experiments'' throughout denotes the mean of a per-experiment statistic
across the ten experiments (reported as mean $\pm$ SEM over experiments).

\paragraph{Trajectory over cycles.} To test whether surfaced theories become
more population- and task-invariant as the loop proceeds, the surviving theories
at the end of each cycle (the un-killed survivor plus that cycle's replacement)
were replayed on all ten experiments and scored as above; the round-0 seeds were
evaluated as cycle-invariant reference lines (Fig.~\ref{fig:human1}C,D;
Fig.~\ref{fig:human2}C,D). We tested the cycle effect with a linear mixed-effects
model regressing per-experiment $MSE_{\hat p(B)}$ on cycle index with a random
intercept per experiment, reporting the fixed-effect slope $\beta$ and its Wald
$z$/$p$. Theories were compared to one another with two-sided paired $t$-tests
across the ten experiments (effect size Cohen's $d_z$).

\paragraph{Held-out generalization for binary-featured decision-making task.} Generalization was scored on
\cite{hilbig2014generalized} (Exp.~1) with that study's design pinned (its fixed
validities, $r_{\max}=1$, and feature count) regardless of the run's own designs;
each surfaced theory and the canonical baselines (TTB, Tallying, WADD) were
replayed under the identical $N=200$ seeded procedure and compared to the
human $\hat p(B)$ via $MSE_{\hat p(B)}$ and $r_{\hat p(B)}$
(Fig.~\ref{fig:human1}G--J).

\subsection*{Preregistered follow-up study}

We conducted a preregistered follow-up study to test predictions derived from Diminishing Returns WADD, the strongest theory surfaced by \textsc{AutoCog}. The preregistration was submitted before data collection and is available at \href{osf-preregistration}{https://osf.io/f7kes}. The full preregistration contains the frozen stimulus battery, model definitions, power simulations, exclusion criteria, and confirmatory analysis code.

The follow-up study consisted of two separate between-subjects online experiments. In both experiments, participants chose between two fictitious products described by four numerical feature values on a common 0--5 scale. Each feature was accompanied by a displayed expert accuracy, indicating how strongly that feature should count toward the choice. Participants completed 96 trials. Trial order and option side were randomized within participants, and participants were not told the hypotheses or model predictions.

Experiment 1 was a model-discrimination experiment. It included stimuli designed to separate the predictions of Diminishing Returns WADD from three alternative decision rules: linear WADD, tallying, and take-the-best. The stimulus battery was divided into three subsets, one for each comparison. Within each subset, Diminishing Returns WADD and the corresponding alternative model made opposing predictions.

Experiment 2 was a value-curvature experiment. It included two types of trials. Steep-vs-flat trials contrasted equal-sized value advantages occurring in different parts of the value scale: the lower, steeper region versus the higher, flatter region. Level-shift trials presented the same feature-value structure once in a low value range and once shifted upward into a higher value range. These trials were designed to test whether the same objective differences had weaker effects when they occurred higher on the value scale.

Sample sizes were fixed before data collection using the power simulation included in the preregistration. Experiment 1 recruited 50 completed participants, balanced across five expert-accuracy vectors. Experiment 2 recruited 100 completed participants, also balanced across the same five vectors. The confirmatory analyses were conducted on participant-level choice proportions, with all tests preregistered as one-tailed tests in the predicted direction.

\subsection*{Controls}

\paragraph{Neutral framing versus adversarial framing in experimental design:} As a control for adversarial framing during experimental design, we considered an experimental-design agent with neutral framing. 
In this framing, both theories are presented symmetrically — neither advocated nor competing — and the agent is prompted to ``design an experiment whose data will best distinguish the two theories — you have no stake in either theory''. 
We compared the two framings in a simulation setting aimed at recovering three ground-truth theories (TTB, take-the-worst (TTW), and perseveration) on two measures: (1) how well each proposed design separates the two candidate theories, quantified as the excess sequence-JSD of the first proposed design over a size-matched random-design baseline; and (2) whether the final surfaced theory recovered the ground-truth behavior and mechanism. 
With three runs per condition, both framing produced designs that separated the models well above the random baseline, with no detectable difference between them: TTW (neutral 0.091 $\pm$  0.016 vs. adversarial 0.095 $\pm$ 0.040), TTB-sampling (0.120 $\pm$  0.061 vs. 0.067 $\pm$  0.020), and perseveration (0.097 $\pm$  0.030 vs. 0.076 $\pm$ 0.029; mean $\pm$ sd). The final recovery was also equivalent for TTB and TTW. Perseveration was the exception: adversarial framing recovered perseverative behavior in 3/3 runs (as early as the second cycle) versus 0/3 for neutral, and only adversarial framing surfaced theories whose code actually reads choice history (2/3 vs. 0/3).

\paragraph{Hard-coded metric versus LLM-proposed metric:}  As a control for the LLM-proposed metric (baseline), we replaced it with an explicit distance: Jensen-Shannon divergence (JSD) between the models' conditional choice distributions given the current stimulus and the previous response (lag-1). 
Assume that an experiment $E$ with $T$ trials is presented in a single order shared among the simulated participants.  For theory $\theta$ we draw $N$ synthetic participants, participant $n$ with parameters $\phi_n \sim p_\theta(\phi)$, and record the binary choice sequence $r^{\theta}_{n} \in \{0,1\}^{T}$; stacking participants gives a response matrix $R^{\theta}\in\{0,1\}^{N\times T}$. Pooling over
participants marginalises each theory's parameters and policy stochasticity, so the resulting frequencies are plug-in estimates of $\theta$'s posterior-predictive choice distribution. At each trial $t\ge 2$ we form the empirical joint over the pair of consecutive responses $(r_{t-1}, r_t)$,
\begin{equation}
  \widehat{P}^{\theta}_{t}(i,j)
  \;=\; \frac{1}{N}\sum_{n=1}^{N}
        \mathbf{1}\!\left[r^{\theta}_{n,t-1}=i \,\wedge\, r^{\theta}_{n,t}=j\right],
  \qquad (i,j)\in\{0,1\}^2,
\end{equation}
a distribution over four cells $\{00,01,10,11\}$; the first trial, which has no predecessor, contributes the marginal $\widehat{P}^{\theta}_{1}(i)=\frac1N\sum_n \mathbf{1}[r^{\theta}_{n,1}=i]$. For two distributions $P,Q$ on the same support the Jensen--Shannon divergence (in nats, bounded by $\ln 2$) is
\begin{equation}
  \mathrm{JSD}(P,Q)
  \;=\; \tfrac12\,\mathrm{KL}\!\left(P \,\|\, \bar M\right)
       + \tfrac12\,\mathrm{KL}\!\left(Q \,\|\, \bar M\right),
  \qquad \bar M=\tfrac12(P+Q),
\end{equation}
with $\mathrm{KL}(A\|B)=\sum_x A(x)\log\frac{A(x)}{B(x)}$ and the convention
$0\log 0 = 0$. The discriminability lag-1 between two theories of design $E$ is the mean between trials of the JSD per-trial between their consecutive-choice distributions,
\begin{equation}
  D^{(1)}(\theta_1,\theta_2; E)
  \;=\; \frac{1}{T}\sum_{t=1}^{T}
        \mathrm{JSD}\!\left(\widehat{P}^{\theta_1}_{t},\,
                            \widehat{P}^{\theta_2}_{t}\right),
\end{equation}
and the design is accepted iff $D^{(1)}(\theta_1,\theta_2; E) > \tau$. 
We use $N=300$ and $\tau=0.041$; because $\widehat{P}^{\theta}_{t}$ is a finite-sample plug-in estimate, $D^{(1)}$ is upward biased, so $\tau$ is calibrated at the same $N=300$ and the bias cancels in the comparison.

The baseline accepts a proposed design when an LLM-proposed statistic separates the two theories under a Welch test at the
planned human sample size. We instead simulate $n{=}300$ synthetic
participants from each theory on the design and accept it iff the lag-1 JSD between the two theories' choice sequences exceeds a pre-registered threshold $\tau$: for each trial position $t$ we form the $2{\times}2$ joint over consecutive responses $(r_{t-1}, r_t)$ across simulated runs, take the JSD between the two theories' joints, and average over positions. The threshold $\tau = 0.041$ is the median of this quantity over $50$ random designs for the canonical theory pairs, fixed before running the control; calibration and runtime both use $N{=}300$.
Regarding theory analysis, the baseline evaluates its accepted metric
$m(\cdot)$ on the simulated data from ground truth theories and on each theory's simulations. We retain this plumbing and only redefine $m$: the metric of a dataset is the lag-1 JSD between its conditional choice distribution $\hat P(r_t \mid \text{stimulus},\, r_{t-1})$ and that of the design's proposing theory. 

We compared the two metrics on the task of recovering three synthetic ground truths: TTB, TTW, and Perseveration, with three runs each and otherwise identical settings to \textsc{AutoCog} baseline. 
Recovery was scored by replaying the surfaced theory on Hilbig (2014) stimuli and comparing it to the three noiseless ground truth theories, in terms of recovery of its underlying mechanism. 
We found that \textsc{AutoCog} was able to recover TTB (3/3 runs each) irrespective of the metric. However, \textsc{AutoCog} with JSD failed to recover perseveration (0/3 runs; one run produced only a cue-following model with an added choice-inertia term) and take-the-worst (0/3), whereas \textsc{AutoCog} with LLM-proposed metrics recovered perseveration in 3/3 runs and take-the-worst in 2/3 (once exactly, once as a mixture component). 
Beyond performance, fixing an explicit distance fixes in advance which behavioral differences count as evidence: theories that differ only in ways the chosen distribution cannot express are indistinguishable to the pipeline, and therefore earn no selection pressure. 
As trial-independent choice of distribution is blind to sequential structure (e.g., perseveration and alternation share near-identical per-trial choice probabilities), we consider lag-1 conditioning to capture first-order dependence. 
Extending the distance to richer joints over response sequences gets computationally expensive (resource and execution time wise) fairly quickly as models and data scale, and in addition requires enforcing stringent constraints on the experimental design space (e.g., match trial sequence across participants that prevents us from extracting rich subject-level structure).
The LLM-proposed metric sidesteps this rigidity: each proposed statistic is an adaptive, interpretable contrast tailored to the current theory pair, evaluated per subject so that between-subject variance — which proved diagnostic for perseveration — enters the loop as evidence.

\paragraph{Random experimental design versus experimental design proposed by LLM:} To test the importance of sampling experiments with LLMs, we replaced it with programmatically generated random experimental designs while keeping the rest of the pipeline the same, including the LLM-based metrics. 
Concretely, in place of the two adversarial proposers, we drew each cycle's designs from a random sampler that returns a valid decision-making experiment: random cue validities (with an enforced spread, to avoid degenerate uniform-validity designs) and random, non-identical binary cue profiles for the two options. Both proposer slots used this sampler, so each cycle still produced two designs which match the proposal budget and the cycle structure of the adversarial baseline. A design was accepted whenever it was valid and both candidate theories could be simulated on it, dropping only the adversarial discriminability test (which is ill-defined for designs that are not optimized to separate the theories).
We recovered three synthetic ground truths: TTB, TTW, and Perseveration, with three runs each and otherwise identical settings to the \textsc{AutoCog} baseline. Recovery was scored by replaying the surfaced theory on the stimuli from \cite{hilbig2014generalized} and comparing it to the noise-free ground truth in terms of recovery of its underlying mechanism. 
We found that \textsc{AutoCog} was able to recover TTB (3/3 runs each) irrespective of the experimental type of experimental design. But \textsc{AutoCog} with random design failed to recover perseveration (0/3 runs) and take-the-worst (0/3), whereas \textsc{AutoCog} with LLM-proposed experimental design recovered perseveration as a mechanism in 3/3 runs and take-the-worst in 2/3 runs. 
In addition, as the proposed LLM metric carries information about the theories being distinguished and has flexibility to adapt to any instance of the experiments, we ran \textsc{AutoCog} with random experimental design paired with JSD between model conditionals as well as control.  We found that this version of \textsc{AutoCog} was only able to recover TTB (1/3 runs each). perseveration (0/3 runs) and take-the-worst (0/3).
Together, these results indicate that the LLM-based experimental design leads to better theoretical recovery compared to the random experimental design, even more so when paired with a theory-agnostic metric. 





\subsection*{Data, software, and code}
Data, code, and analysis scripts are available at \href{github.com/akjagadish/autocog/}{https://github.com/akjagadish/autocog/}\\
Parts of the text were refined with the help of generative AI tools, such as ChatGPT, Claude, and Writefull, which provided suggestions for rewording, paraphrasing, and restructuring. All outputs were carefully reviewed and edited prior to use.

\section*{Acknowledgements}

This work was supported by funding from the Eric and Wendy Schmidt Transformative Technology Fund at Princeton University. AKJ is supported by a Natural and Artificial Mind (NAM) Fellowship from the Scully Peretsman foundation.


\section*{Author contributions statement}

AKJ, YS, and SHC conceived the study;
AKJ, YS, and SHC developed the methodology and theoretical framework; 
NJ, ES, TLG, GK and ND contributed to the refining of the theoretical framework;
AKJ and YS designed and conducted the experiments with input from SHC; 
AKJ and YS collected and preprocessed the data and generated the figures; 
AKJ, YS, and SHC wrote the initial draft of the manuscript;
Everyone reviewed and edited the manuscript; 
TLG acquired funding.

\bibliography{main}

\clearpage
\newpage

\begin{center}
    \LARGE \textbf{Supplementary Information}\\[1em]
\end{center}
\setcounter{figure}{0}
\renewcommand{\figurename}{Figure}
\renewcommand{\thefigure}{S\arabic{figure}}
\renewcommand{\tablename}{Table}
\renewcommand{\thetable}{S\arabic{table}}

\section*{The \textsc{AutoCog} cycle in detail as figures}
\label{si:cycle}


\begin{figure}[!h]
\centering
\includegraphics[width=\linewidth]{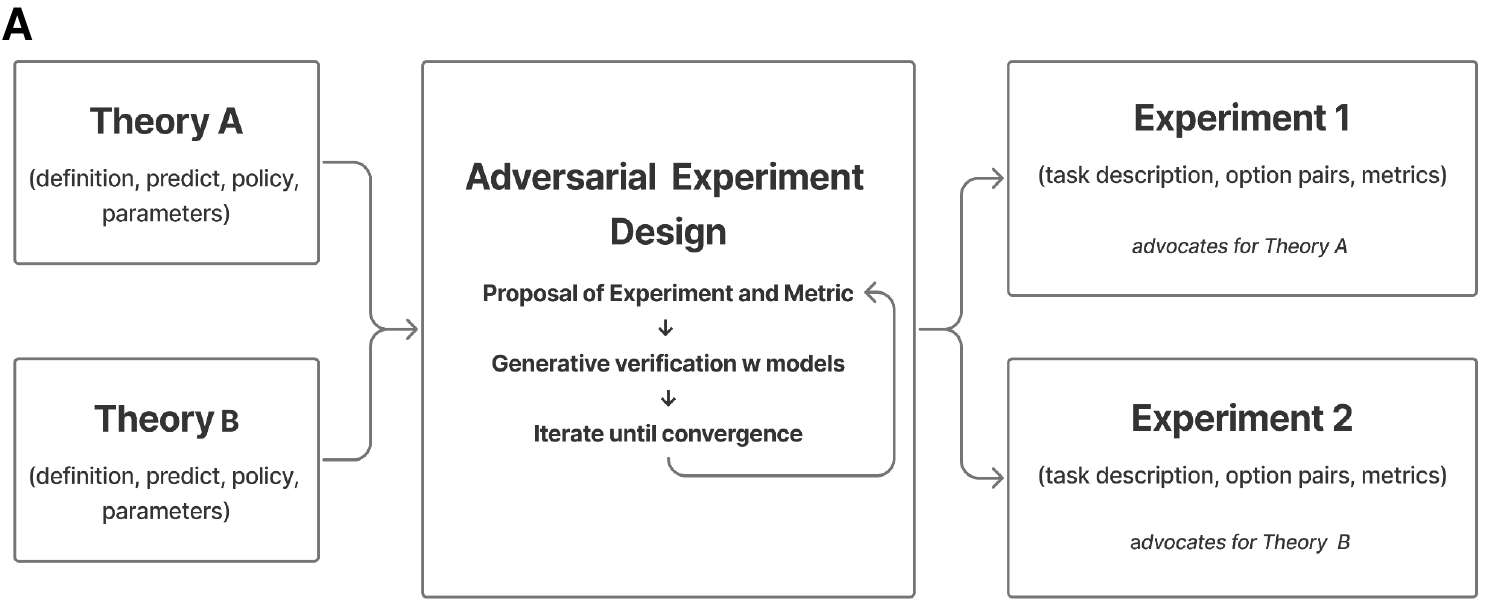}
\caption{\textbf{Adversarial experimental design (stage 1).} The two competing
theories, each an executable object (a natural-language \emph{definition}, a
\texttt{predict} function, a \texttt{policy} function, and \emph{parameters}
with admissible ranges), initialize two proposer agents. Each agent proposes
an experiment (task description, option pairs, metric) under which its own
theory should outperform its competitor, and iterates proposal and generative
verification until convergence. A pair is accepted only once verification
confirms that the metric separates the two theories at the planned sample
size, yielding one experiment per theory.}
\label{fig:si-expdesign}
\end{figure}


\begin{figure}
\centering
\includegraphics[width=\linewidth]{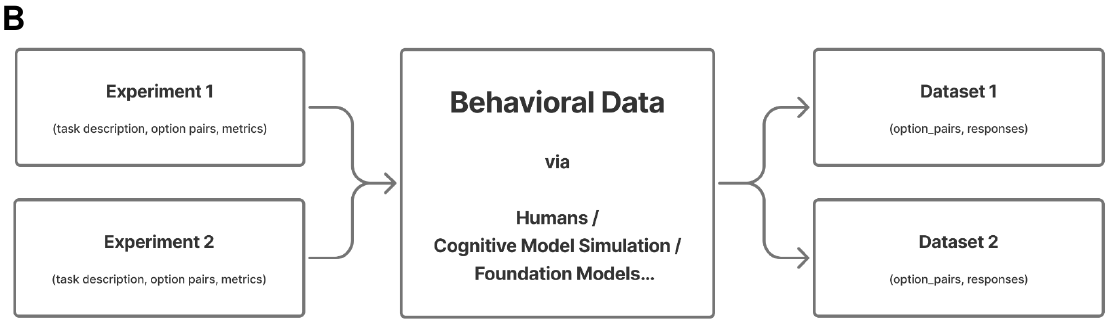}
\caption{\textbf{Behavioral data collection (stage 2).} Each accepted
experiment is administered to the run's behavior source which can be either human participants recruited online, or simulations from a cognitive model or a behavioral foundation model. The same pipeline applies across all three. Each experiment returns a dataset of option pairs and observed responses that enters the shared observation pool.}
\label{fig:si-datacollection}
\end{figure}


\begin{figure}
\centering
\includegraphics[width=\linewidth]{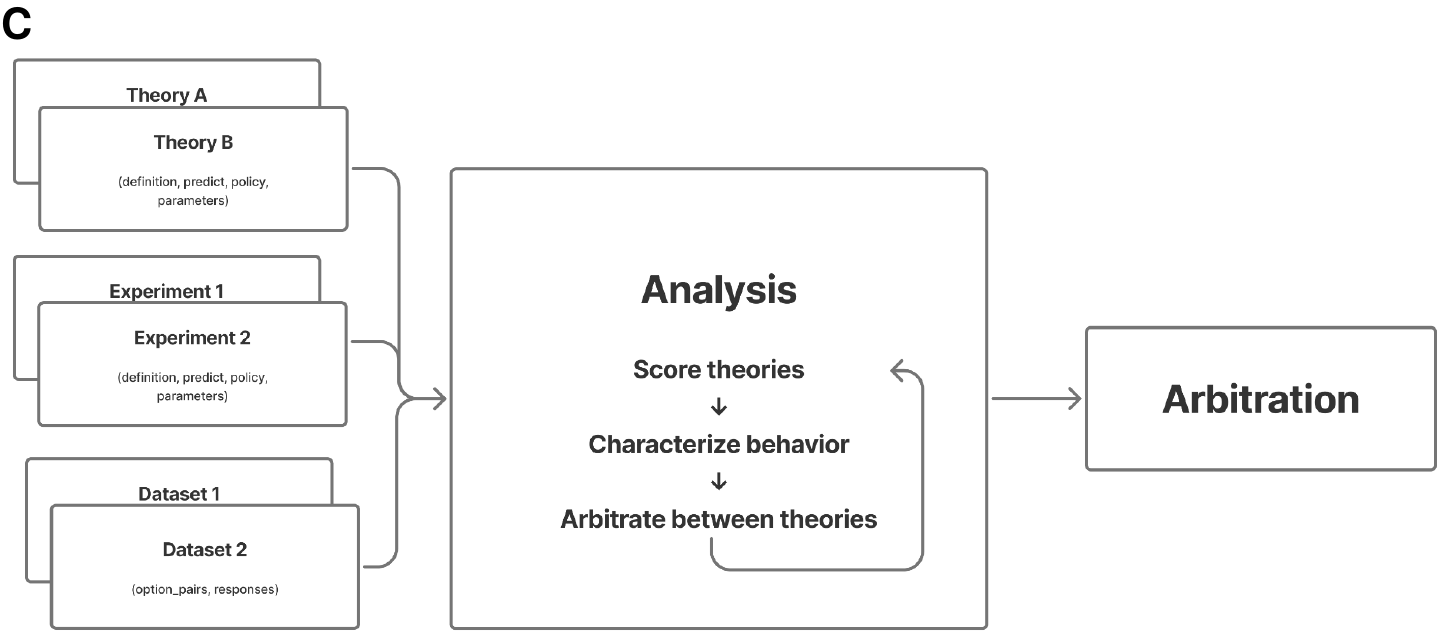}
\caption{\textbf{Analysis and arbitration (stages 3--4).} All theories,
experiments, and datasets seen so far enter an analysis loop that scores every
theory on every observation by forward simulation, characterizes the observed
behavior, and ranks theories on a shared leaderboard. A neutral arbitrator
then reviews the cycle and returns a structured verdict.}
\label{fig:si-analysis}
\end{figure}


\begin{figure}[t]
\centering
\includegraphics[width=\linewidth]{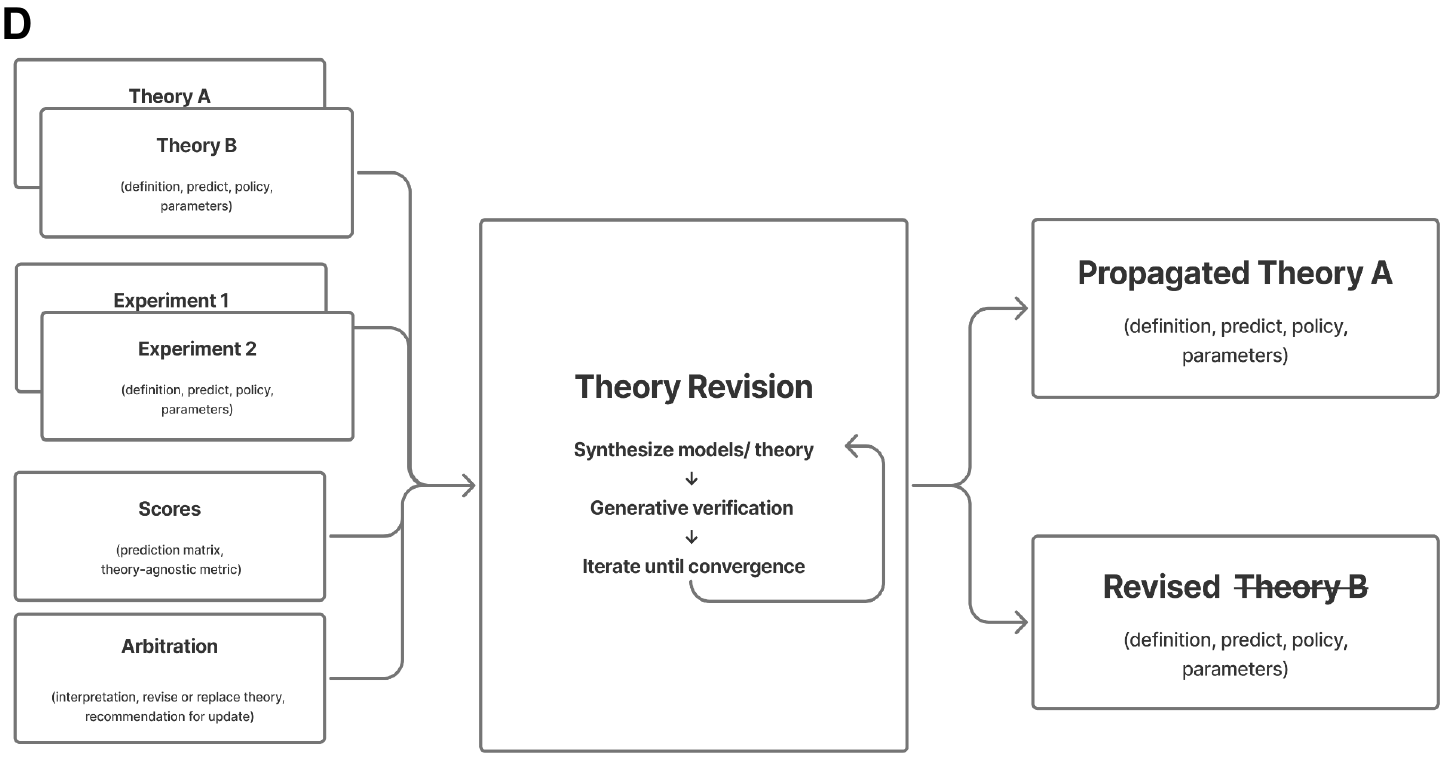}
\caption{\textbf{Theory revision (stage 5).} Conditioned on the scores and the
arbitrator's verdict, the winning slot is propagated unchanged while the
losing slot is revised (its executable components only) or replaced (its
description and model). Candidates are synthesized in an inner critique loop
and accepted only if they improve aggregate fit on the full observation pool,
mirroring the generative verification used in stage 1. The revised theory
seeds the next cycle.}
\label{fig:si-revision}
\end{figure}


\clearpage
\section*{Baseline theories}
\subsection*{Canonical theories}
\label{si:canonical-theories}
The three canonical decision-making heuristics span the classical
frugality--compensation spectrum: a one-reason lexicographic rule (Take The
Best), an unweighted vote (Tallying), and a fully compensatory weighted sum
(Weighted Additive). As above, response noise enters through a softmax over the
choice score with inverse temperature $\beta$, plus an independent lapse that
with probability $\epsilon$ replaces the softmax output with a uniform guess over
the two options.

\paragraph{Take The Best \citep{gigerenzer1996reasoning}.}
People compare two options by consulting cues one at a time in order of
validity, stopping at the first cue that discriminates between the two options.
That cue alone determines the choice: the option with the higher value on the
discriminating cue wins, and no other cue is consulted. Lower-validity cues are
never reached once a higher-validity cue discriminates, so Take The Best is a
``one-reason'' decision rule---only a single feature is ever used on any given
choice. Because only the sign of the comparison on the top discriminating cue
matters, it ignores both the magnitude of that difference and all
lower-validity information, making it maximally frugal in its use of evidence.
Cue validities are subjective and may be learned, inferred from instructions, or
fixed by task structure. When no cue discriminates, the learner must guess.

\paragraph{Tallying \citep{dawes1979robust, gigerenzer1999simple}.}
People compare two options by counting, across all features, how often one
option has a higher value than the other; the option that wins on more features
is chosen. Tallying discards cardinal magnitudes---only the sign of each
feature-wise comparison matters---so it is robust to monotone rescaling of
individual features and cannot be swayed by a single large feature difference.
Ties on an individual feature contribute to neither count. Unlike Take The Best,
no feature is privileged: every cue contributes equally to the tally. When the
two counts are equal the heuristic has no basis for preference and the learner
must guess. Here the softmax operates over the two tallies, interpolating
between deterministic choice at large $\beta$ and uniform guessing at $\beta=0$.

\paragraph{Weighted Additive \citep{payne1993adaptive}.}
People compute, for each option, a weighted sum of its feature values, with each
feature weighted by its subjective validity (or importance); the option with the
higher weighted sum is chosen. WADD is the compensatory benchmark against which
one-reason heuristics are contrasted: a large deficit on a high-validity cue can
be compensated by a sufficiently strong advantage on lower-validity cues, so no
single feature is ever decisive on its own. Unlike Tallying, WADD uses cardinal
feature magnitudes and weights them by validity, exploiting both the sign and
the size of each comparison as well as inter-cue differences in informativeness;
unlike Equal-Weight, its weights differ across features. Behavior is invariant to
a shared affine rescaling across options but scales linearly with per-feature
weight. When the two weighted sums are equal the learner must guess.

\subsection*{Non-canonical theories}
\label{si:noncanonical-theories}

To stress-test recovery beyond the three canonical heuristics (Take The Best,
Tallying, and Weighted Additive), we additionally authored six non-canonical
decision-making theories. These were chosen to span distinct failure modes for
a recovery procedure: inverted cue cascades, extreme frugality, sequential
dependencies that ignore the stimulus, and contrarian ensembles. Unless noted, response noise enters through a softmax over the binary choice score (winner $=1$, loser $=0$) with inverse temperature $\beta$, plus an independent lapse that with probability
$\epsilon$ replaces the softmax output with a uniform guess over the two
options.

\paragraph{Take The Worst (inverted Take The Best).}
People compare two options by consulting cues one at a time, but in the
\emph{wrong} order---from least valid to most valid---stopping at the first cue
that discriminates between the two options. That cue alone determines the
choice. This is the adversarial inverse of Take The Best: it preserves TTB's
one-reason, lexicographic, maximally frugal character but follows an
\emph{ascending}-validity cue cascade rather than a descending one. The result
is a model that is perfectly predictive of choices yet leans almost entirely on
the low-validity features the design does not bias toward. When two options
agree on the least-valid cue the cascade climbs to the next-least-valid cue, and
so on; only when every cue ties must the learner guess.

\paragraph{Single-Cue (extreme one-reason frugality).}
People base every choice on a \emph{single} cue---the one with the lowest
validity---and ignore all other cues entirely. The option scoring higher on
that one least-valid cue is chosen; if the two options tie on it, the learner
consults nothing else and simply guesses. Unlike Take The Best (or Take The
Worst) there is no cue cascade: even when the least-valid cue fails to
discriminate, the model never falls back to another cue. Exactly one feature is
ever read, fixed in advance as the least-valid expert, making this strictly more
frugal than any lexicographic rule.

\paragraph{Anti-Majority Ensemble (contrarian vote).}
People are modeled as running the three classic heuristics in
parallel---Take The Best, Tallying, and Weighted Additive---reading off the
single option each heuristic deterministically prefers, and then leaning toward
whichever option the majority of the three did \emph{not} prefer. With three
binary voters at least two always agree, so a majority option always exists; the
model targets its complement. The individual heuristics contribute only their
hard preference (their argmax) to the vote and carry no internal noise, so
$\beta$ and $\epsilon$ describe noise in the contrarian response itself.

\paragraph{Perseveration (pure repeat).}
People ignore the stimulus entirely and simply repeat their previous response
from trial to trial. The first choice has no prior to repeat, so it is made at
random; on every subsequent trial the response is identical to the previous
response. This is a pure sequential-dependency baseline that depends only on the
choice history, not on any feature ratings or validities. It is the exact mirror
image of the alternation baseline---the two are distinguished purely by the sign
of their lag-1 sequential dependency ($+1$ for perseveration, $-1$ for
alternation).

\paragraph{Alternation (pure switch).}
People ignore the stimulus entirely and simply alternate their response from
trial to trial. The first choice is made at random; on every subsequent trial
the response is the opposite of the previous one. Like perseveration, this is a
sequential-dependency baseline that depends only on the choice history,
capturing a subject who ``takes turns'' between the options regardless of
evidence.
\section*{Hyperparameters and budgets}
\label{si:hyperparams}

Table~\ref{tab:si-hyperparams} lists the fixed settings used across runs
unless stated otherwise.

\begin{table}[h]
\centering
\caption{\textsc{AutoCog} settings used across runs.}
\label{tab:si-hyperparams}
\begin{tabular}{lll}
\toprule
Stage & Setting & Value \\
\midrule
Run            & Cycles per run                       & 5 (20 for harder non-canonical strategies) \\
Run            & Theory slots                         & 2 \\
Design (1)     & Experiments attempted per agent      & $\leq 3$ \\
Design (1)     & Metrics attempted per experiment     & $\leq 4$ \\
Design (1)     & Discriminability test                & Welch two-sample $t$-test, two-sided \\
Design (1)     & Acceptance threshold                 & $\alpha = 0.01$ \\
Design (1)     & Planned sample size for test         & $N = 25$ \\
Data (2)       & Participants per experiment          & $N = 25$ \\
Data (2)       & Response exclusion                   & RT $< 1500$\,ms \\
Data (2)       & Median completion / compensation     & ${\sim}8$\,min / US\$0.80 \\
Revision (5)   & Critique-loop iterations             & $\leq 10$ \\
Revision (5)   & Compilation/validation retries       & $\leq 10$ \\
Models         & Language model                       & \textsc{Gemini-3.1-Pro} \\
Models         & Max output tokens                    & 32768 \\
\bottomrule
\end{tabular}
\end{table}


\section*{Design choices and justifications}
\label{si:design}

\textsc{AutoCog} was built around five design commitments.
\textbf{(1) Natively agentic.} Every generative step (experiment proposal,
metric proposal, arbitration, theory and model synthesis) is delegated to a
language-model agent, so the system improves as agentic harnesses improve and
no stage hard-codes a fixed search space of theories.
\textbf{(2) Minimal but essential human intervention.} A researcher specifies
only the experimental design space, the model design space, the seed theories, and
the analysis metric; the loop then runs without further intervention, while
still allowing researchers to narrow the search by pointing to factors of
interest.
\textbf{(3) Iterative self-verification.} Both the experiment-design stage and
the revision stage follow a propose--verify--update procedure, in which a
proposal is accepted only after it is checked against simulated or observed
behavior, rather than trusting a single language-model output.
\textbf{(4) Fast and scalable.} Theories are evaluated by forward simulation
rather than by fitting each theory to each dataset. Generative evaluation is
faster at scale, is harder to game than a single predictive score, and
implicitly regularizes against overfitting, because a theory must reproduce
the observed distribution of behavior when simulated, not merely fit it after
the fact.
\textbf{(5) Interpretable.} The full reasoning trajectory for each proposed
experiment, metric, verdict, and theory is logged, so a run is an auditable
artifact from which another researcher or agent can resume the search.

\input{si_prompts.tex}
\label{si:prompts}

\subsection*{LLM-as-judge rubric}
\label{si:judge-rubric}

To quantify how well a surfaced theory recovers the \emph{mechanism} of the
data-generating ground truth (independent of behavioral fit), we scored each
(surfaced, ground-truth) pair with an LLM judge
(\textsc{Gemini gemini-3-flash-preview}). The judge rates the similarity of
the two models' underlying decision rule on a continuous $[0,1]$ scale,
explicitly disregarding wording, terminology, framing, and parameter values.
Output is enforced as strict JSON
(\texttt{\{"similarity": float, "rationale": str\}}). Each pair is scored under
two input modes --- \emph{description} (natural-language descriptions only) and
\emph{joint} (descriptions plus each theory's \texttt{predict()} source) --- and,
to reduce sampling variance, scored three times per mode with the scores averaged.

\paragraph{Judge system prompt (verbatim).}
\begin{Verbatim}[breaklines=true,breakanywhere=true,fontsize=\footnotesize]
You are a cognitive scientist comparing two cognitive models in a decision-making task. Rate how similar their UNDERLYING DECISION MECHANISM is -- the rule each model uses to turn inputs into a choice -- ignoring wording, terminology, framing, and parameter values. Score in [0, 1]: 1.0 = the same decision rule (they would make the same choices for the same reason); 0.5 = a recognizable variant or partial overlap (shares core structure but deviates non-trivially); 0.0 = an unrelated mechanism (a genuinely different rule). Use the full range and intermediate values; be calibrated, not generous. Reply with strict JSON matching {"similarity": float, "rationale": str}; rationale is one short sentence (<= 50 words).
\end{Verbatim}

\paragraph{Judge user prompt (structure).}
The user message presents the two theories back to back, separated by a
horizontal rule. Both the description blocks and the
fenced \texttt{predict()} source are used.
\begin{Verbatim}[breaklines=true,breakanywhere=true,fontsize=\footnotesize]
# Ground-truth theory: {gt_name}
**Description:**
{gt_description}

**predict():**          
```python
{gt_predict_source}
```

---

# Surfaced theory: {surfaced_label}
**Description:**
{surfaced_description}

**predict():**         
```python
{surfaced_predict_source}
```
\end{Verbatim}

\paragraph{Rubric anchors.}
\begin{itemize}
\item \textbf{1.0} --- the same decision rule (identical choices, for the same reason).
\item \textbf{0.5} --- a recognizable variant or partial overlap (shared core
structure with non-trivial deviation).
\item \textbf{0.0} --- an unrelated mechanism (a genuinely different rule).
\end{itemize}
The judge is instructed to use the full range and intermediate values and to be
calibrated rather than generous. The reported similarity for a (surfaced,
ground-truth) pair is the arithmetic mean of the three sampled scores; the
first sample's one-sentence rationale is retained for auditing. Because
$\epsilon=1.0$ corresponds to a fully randomized ground truth (recovery is
undefined), it is excluded from the reported figures.

\section*{Human experiments}
\label{si:human}

\paragraph{Platform stack.}
Each committed experiment was compiled to a browser task with
SweetBean~\cite{sweetbean} and deployed through AutoRA~\cite{Musslick_AutoRA_Automated_Research_2024}.
Recruitment ran on Prolific~\cite{palan2018prolific} while the task itself was
served as a hosted web app and all trial-by-trial responses were written to a
Google Firebase (Firestore) backend; AutoRA's Firebase--Prolific runner
coordinates the two, publishing one Prolific study per committed experiment,
collecting completed sessions, and returning the response data to the loop.
Within a cycle the two committed experiments (one advocated by each theory) were
launched as separate conditions, so $2N$ participants were recruited per cycle.

\paragraph{Consent and ethics.}
The studies were conducted under a Princeton University IRB protocol (study
title: ``Towards a unified understanding and reliable estimation of cognitive
processes''). Eligible participants were adults aged 18--55. Before the task,
participants were shown an informed-consent screen describing the purpose
(human performance in simple computerized tasks), the minimal-risk nature of
participation (possible mild fatigue or boredom), voluntary participation with
the right to withdraw at any time, and the data-handling procedures; they
proceeded only after giving consent.

\paragraph{Recruitment and payment.}
A single completion code returned
participants to Prolific on finishing the task. The per-condition session was
capped at a 20-minute timeout. Median completion was approximately 6--8 minutes,
and participants were compensated US\$0.80 for the session.

\paragraph{Trial-level response gate.}
To ensure participants inspected the full stimulus before responding, the
response keys were locked and the ``A''/``B'' prompt hidden for the first
1500\,ms of every trial (implemented per stimulus via SweetBean's
\texttt{HtmlKeyboardResponse(min\_rt=1500)}). This is a per-trial pacing gate,
not a data-exclusion criterion: responses are simply not accepted until the
window elapses.

\begin{table}[t]
\centering
\caption{Preregistered follow-up tests of predictions generated by Diminishing Returns WADD.}
\label{tab:preregistered-followup}
\begin{threeparttable}
\begin{tabular}{
  l
  l
  c
  S[table-format=1.3]
  S[table-format=1.3]
  c
  c
  S[table-format=<1.2e-1]
}
\toprule
{Hypothesis} & {Test} & {Match / rate} & {$M$ or $M_{\Delta}$} & {SE} & {95\% CI} & {$t(\mathrm{df})$} & {$p$} \\
\midrule
H1 & Concave vs.\ WADD     & .632 vs.\ .368 & 0.264 & 0.076 & [0.111, 0.417]  & 3.465 (49) & {$5.56\times10^{-4}$} \\
H1 & Concave vs.\ Tallying & .678 vs.\ .322 & 0.356 & 0.063 & [0.230, 0.483]  & 5.665 (49) & {$3.83\times10^{-7}$} \\
H1 & Concave vs.\ TTB      & .634 vs.\ .366 & 0.268 & 0.068 & [0.131, 0.404]  & 3.927 (49) & {$1.35\times10^{-4}$} \\
H2 & Steep-vs-flat         & .582           & 0.582 & 0.020 & [0.543, 0.621]  & 4.192 (99) & {$3.01\times10^{-5}$} \\
H3 & Level-shift           & .774 vs.\ .745 & 0.029 & 0.016 & [-0.003, 0.061] & 1.822 (99) & {.036} \\
\bottomrule
\end{tabular}
\begin{tablenotes}
\footnotesize
\item Diminishing Returns WADD was tested against each alternative model with one-sided paired (H1) or one-sample (H2, H3) $t$-tests in the predicted direction; H1 comparisons were Holm-corrected. $N = 50$ participants for H1 ($\mathrm{df} = 49$) and $N = 100$ for H2 and H3 ($\mathrm{df} = 99$); SE denotes the standard error of the mean. For H1, $M_{\Delta}$ is the difference between the Diminishing Returns WADD match rate and the alternative-model match rate. For H2, $M$ is the proportion of steep-vs-flat trials on which participants chose the low-range advantage option. For H3, $M_{\Delta}$ is the difference between target-option choice rates in the low-range and shifted-up versions. The 95\% CIs are two-sided; consequently the H3 interval includes zero even though the preregistered one-sided test is significant at $p = .036$.
\end{tablenotes}
\end{threeparttable}
\end{table}


\section*{Surfaced cognitive models from the human-study runs}
\label{si:surfaced-code}

For each human study we reproduce the two theories surviving the final cycle, exactly as synthesized by \textsc{AutoCog}: the natural-language description, the parameter ranges, and the executable \texttt{predict} (\texttt{generate}) and \texttt{policy} source. The theory reported as the run's winner is marked \textbf{(winning theory)}.

\subsection*{Study 1 — Binary-valued cues (seeds: TTB, WADD)}

\paragraph{Non-linear Subjective Weighting Model \textbf{(winning theory)}}
\textit{Description.} Non-linear Subjective Weighting Model: Subjects evaluate options by computing a weighted sum of their features, but they do not use the objective cue validities directly. Instead, subjective cue weights are a power function of the provided validities. An individual-specific exponent parameter controls the non-linearity of this transformation. This single mechanism unifies multiple decision strategies: an exponent near 0 flattens the weights (yielding Equal-Weight/Tallying), an exponent of 1 uses the validities linearly (yielding WADD), and a large exponent strongly amplifies the most valid cues (yielding non-compensatory Take The Best behavior).

\textit{Parameters.}
\begin{Verbatim}[breaklines=true,breakanywhere=true,fontsize=\footnotesize]
gamma: [0.0, 10.0]
beta: [0.1, 15.0]
epsilon: [0.0, 0.5]
validities: validities
\end{Verbatim}

\textit{predict / generate.}
\begin{Verbatim}[breaklines=true,breakanywhere=true,fontsize=\footnotesize]
def predict(parameters, state, history):
    import numpy as np
    
    stim = np.asarray(state, dtype=float)
    if stim.ndim != 2 or stim.shape[0] != 2:
        raise ValueError(f"Expected a (2, n_features) stimulus; got shape {stim.shape}.")
        
    validities = np.asarray(parameters["validities"], dtype=float)
    gamma = float(parameters["gamma"])
    beta = float(parameters["beta"])
    epsilon = float(parameters["epsilon"])
    
    # Non-linear subjective weighting: w_i = v_i ^ gamma
    # Validities are in [0.5, 1.0], so base is positive.
    w = np.maximum(validities, 0.0) ** gamma
    
    # Option scores are the weighted sum of features
    scores = stim @ w
    
    # Softmax over scores with inverse temperature beta
    z = beta * (scores - np.max(scores))
    e = np.exp(z)
    p_core = e / np.sum(e)
    
    # Add independent lapse noise
    return (1.0 - epsilon) * p_core + epsilon * np.array([0.5, 0.5])
\end{Verbatim}

\textit{policy.}
\begin{Verbatim}[breaklines=true,breakanywhere=true,fontsize=\footnotesize]
def policy(probs):
    import numpy as np
    probs = np.asarray(probs, dtype=np.float64)
    probs /= probs.sum()
    return int(np.random.choice(len(probs), p=probs))
\end{Verbatim}

\paragraph{Probabilistic Strategy Mixture Model with Flexible Compensatory Component}
\textit{Description.} Probabilistic Strategy Mixture Model with Flexible Compensatory Component: Subjects maintain a repertoire of distinct cognitive strategies—a non-compensatory heuristic (Take-The-Best) and a compensatory strategy (Weighted Additive). On any given trial, they probabilistically select which strategy to deploy based on an individual trait parameter. To accurately capture heavily compensatory behavior observed in certain environments, the Weighted Additive strategy uses a non-linear power transformation of the validities, allowing subjects to tune their compensatory integration rather than strictly using raw validities.

\textit{Parameters.}
\begin{Verbatim}[breaklines=true,breakanywhere=true,fontsize=\footnotesize]
p_ttb: [0.0, 1.0]
beta: [0.1, 25.0]
epsilon: [0.0, 0.2]
gamma: [0.1, 5.0]
validities: validities
\end{Verbatim}

\textit{predict / generate.}
\begin{Verbatim}[breaklines=true,breakanywhere=true,fontsize=\footnotesize]
def predict(parameters, state, history):
    import numpy as np
    
    stim = np.asarray(state, dtype=float)
    if stim.ndim != 2 or stim.shape[0] != 2:
        raise ValueError(f"Expected a (2, n_features) stimulus; got shape {stim.shape}.")
        
    validities = np.asarray(parameters["validities"], dtype=float)
    p_ttb = float(parameters["p_ttb"])
    beta = float(parameters["beta"])
    epsilon = float(parameters["epsilon"])
    gamma = float(parameters["gamma"])
    
    # Strategy 1: Flexible Weighted Additive (WADD)
    wadd_weights = validities ** gamma
    scores_wadd = stim @ wadd_weights
    z = beta * (scores_wadd - np.max(scores_wadd))
    e = np.exp(z)
    p_wadd = e / np.sum(e)
    
    # Strategy 2: Take-The-Best (TTB)
    order = np.argsort(-validities)
    p_ttb_strat = np.array([0.5, 0.5])
    for idx in order:
        if stim[0, idx] > stim[1, idx]:
            p_ttb_strat = np.array([1.0, 0.0])
            break
        elif stim[1, idx] > stim[0, idx]:
            p_ttb_strat = np.array([0.0, 1.0])
            break
            
    # Mixture of strategies
    p_core = p_ttb * p_ttb_strat + (1.0 - p_ttb) * p_wadd
    
    # Add independent lapse noise
    return (1.0 - epsilon) * p_core + epsilon * np.array([0.5, 0.5])
\end{Verbatim}

\textit{policy.}
\begin{Verbatim}[breaklines=true,breakanywhere=true,fontsize=\footnotesize]
def policy(probs):
    import numpy as np
    probs = np.asarray(probs, dtype=np.float64)
    probs /= probs.sum()
    return int(np.random.choice(len(probs), p=probs))
\end{Verbatim}

\subsection*{Study 2 — Cardinal-valued cues (seeds: TTB, Tallying)}

\paragraph{Threshold-based Binarization (Satisficing WADD)}
\textit{Description.} Threshold-based Binarization (Satisficing WADD): Decision-makers simplify complex cardinal information by converting continuous or multi-level ratings into binary cues based on a satisficing threshold. A feature is considered satisfactory (1) if its rating meets or exceeds the threshold, and unsatisfactory (0) otherwise. The options are then evaluated by computing the validity-weighted sum of these binarized features (WADD on binary cues). This mechanism naturally explains why extreme cardinal advantages (e.g., 10 vs 5) might be ignored if both options exceed the satisficing threshold, allowing an option with distributed moderate advantages to win against an option with a single extreme advantage.

\textit{Parameters.}
\begin{Verbatim}[breaklines=true,breakanywhere=true,fontsize=\footnotesize]
threshold: [0.0, 10.0]
beta: [0.1, 20.0]
epsilon: [0.0, 0.5]
validities: validities
\end{Verbatim}

\textit{predict / generate.}
\begin{Verbatim}[breaklines=true,breakanywhere=true,fontsize=\footnotesize]
def predict(parameters, stimulus, history):
    import numpy as np
    
    stim = np.asarray(stimulus, dtype=float)
    if stim.ndim != 2 or stim.shape[0] != 2:
        raise ValueError("Expects a (2, n_features) stimulus.")
        
    a, b = stim[0], stim[1]
    val = np.asarray(parameters["validities"], dtype=float)
    
    threshold = float(parameters["threshold"])
    beta = float(parameters["beta"])
    epsilon = float(parameters["epsilon"])
    
    # Normalize validities
    sum_val = np.sum(val)
    if sum_val > 0:
        val = val / sum_val
        
    # Binarize features based on the threshold
    bin_a = (a >= threshold).astype(float)
    bin_b = (b >= threshold).astype(float)
    
    # Compute validity-weighted sum of binarized features
    score_a = np.sum(bin_a * val)
    score_b = np.sum(bin_b * val)
    
    scores = np.array([score_a, score_b])
    
    # Convert to probabilities via softmax with max-subtraction
    z = beta * (scores - np.max(scores))
    e = np.exp(z)
    p_core = e / np.sum(e)
    
    n_opts = p_core.shape[0]
    return (1.0 - epsilon) * p_core + epsilon * (np.ones(n_opts) / n_opts)
\end{Verbatim}

\textit{policy.}
\begin{Verbatim}[breaklines=true,breakanywhere=true,fontsize=\footnotesize]
def policy(probabilities):
    import numpy as np
    probs = np.asarray(probabilities, dtype=np.float64)
    probs /= probs.sum()
    return int(np.random.choice(len(probs), p=probs))
\end{Verbatim}

\paragraph{Diminishing Returns WADD posits that individuals evaluate options by applying a concave utility function to cardinal feature values before weighting them by their cue validities \textbf{(winning theory)}}
\textit{Description.} Diminishing Returns WADD posits that individuals evaluate options by applying a concave utility function to cardinal feature values before weighting them by their cue validities. By compressing large cardinal values via a power law transformation, extreme advantages on a single feature are discounted relative to multiple moderate advantages across several features. Parameterizing the shift applied before the power law allows the model to flexibly smooth the extreme marginal utility near zero, preventing over-sensitivity to small integer differences while maintaining the core concave utility mechanism.

\textit{Parameters.}
\begin{Verbatim}[breaklines=true,breakanywhere=true,fontsize=\footnotesize]
alpha: [0.1, 1.0]
beta: [0.1, 20.0]
epsilon: [0.0, 0.5]
shift: [0.1, 5.0]
validities: validities
\end{Verbatim}

\textit{predict / generate.}
\begin{Verbatim}[breaklines=true,breakanywhere=true,fontsize=\footnotesize]
def predict(parameters, stimulus, history):
    import numpy as np
    
    stim = np.asarray(stimulus, dtype=float)
    if stim.ndim != 2 or stim.shape[0] != 2:
        raise ValueError("Expects a (2, n_features) stimulus.")
        
    a, b = stim[0], stim[1]
    val = np.asarray(parameters["validities"], dtype=float)
    
    alpha = float(parameters["alpha"])
    beta = float(parameters["beta"])
    epsilon = float(parameters["epsilon"])
    shift = float(parameters["shift"])
    
    # Normalize validities to stabilize softmax across different experiments
    sum_val = np.sum(val)
    if sum_val > 0:
        val = val / sum_val
        
    # Apply concave utility transformation (diminishing returns)
    # Shifted by a parameterized value to flexibly smooth infinite marginal utility at zero
    u_a = np.power(a + shift, alpha) - np.power(shift, alpha)
    u_b = np.power(b + shift, alpha) - np.power(shift, alpha)
    
    # Compute validity-weighted sum of transformed features
    score_a = np.sum(u_a * val)
    score_b = np.sum(u_b * val)
    
    scores = np.array([score_a, score_b])
    
    # Convert to probabilities via softmax with max-subtraction
    z = beta * (scores - np.max(scores))
    e = np.exp(z)
    p_core = e / np.sum(e)
    
    n_opts = p_core.shape[0]
    return (1.0 - epsilon) * p_core + epsilon * (np.ones(n_opts) / n_opts)
\end{Verbatim}

\textit{policy.}
\begin{Verbatim}[breaklines=true,breakanywhere=true,fontsize=\footnotesize]
def policy(probabilities):
    import numpy as np
    probs = np.asarray(probabilities, dtype=np.float64)
    probs /= probs.sum()
    return int(np.random.choice(len(probs), p=probs))
\end{Verbatim}


\subsection*{How action noise degrades the recovered mechanism}
\label{si:noise-degradation}

To illustrate how recoverability erodes as the data-generating process is corrupted, we fix the ground truth to the sampling variant of Take-The-Best (TTB) and follow a single representative surfaced theory at each action-noise level $\varepsilon\in\{0,0.5,0.75,1.0\}$. As $\varepsilon$ grows, the surfaced mechanism moves from an exact reconstruction of TTB, through TTB variants that try to capture noisy responses with an inflated lapse rate and a softened stopping rule, to a model that abandons cue-based search for near-pure guessing once the ground truth is fully random ($\varepsilon=1.0$). 
The mechanism similarity with the clean TTB ground truth predicted by the LLM-judge falls from $1.00$ to $0.17$ across the sequence. 

\begin{center}
\begin{tabular}{lllll}
\hline
GT noise $\varepsilon$ & Surfaced mechanism & Judge sim. & Lapse $\epsilon$ \\
\hline
0.0 & Take-The-Best (exact) & 1.00 & [0.0, 0.5] \\
0.5 & Take-The-Best with inflated lapse & 1.00 & [0.0, 1.0] \\
0.75 & Take-The-Best with probabilistic stopping & 0.83 & [0.0, 1.0] \\
1.0 & Random Choice / Minimal Effort & 0.17 & [0.9, 1.0] \\
\hline
\end{tabular}
\end{center}

\paragraph{$\varepsilon=0.0$ --- Take-The-Best (exact) (run1, pi\_3; sim $=1.00$).}
Exact deterministic Take-The-Best: lexicographic search, hard stop at the first discriminating cue, choice blended only with a modest uniform lapse.

\begin{Verbatim}[breaklines=true,breakanywhere=true,fontsize=\footnotesize]
# parameters
epsilon: [0.0, 0.5]
validities: validities

def predict(parameters, stimulus, history):
    stim = np.asarray(stimulus, dtype=float)
    if stim.ndim != 2 or stim.shape[0] != 2:
        raise ValueError(f"TTB expects a (2, n_features) stimulus; got shape {stim.shape}.")

    a, b = stim[0], stim[1]
    validities = np.asarray(parameters["validities"], dtype=float)
    
    # Sort features by validity in descending order
    order = np.argsort(validities)[::-1]
    
    chosen = -1
    for idx in order:
        if a[idx] > b[idx]:
            chosen = 0
            break
        elif b[idx] > a[idx]:
            chosen = 1
            break
            
    # Deterministic choice based on the first discriminating cue
    if chosen == 0:
        p_core = np.array([1.0, 0.0])
    elif chosen == 1:
        p_core = np.array([0.0, 1.0])
    else:
        # If all features tie, guess randomly
        p_core = np.array([0.5, 0.5])
        
    epsilon = float(parameters["epsilon"])
    n_opts = p_core.shape[0]
    
    # Blend deterministic choice with uniform lapse rate for noise
    return (1.0 - epsilon) * p_core + epsilon * (np.ones(n_opts) / n_opts)
\end{Verbatim}

\paragraph{$\varepsilon=0.5$ --- Take-The-Best with inflated lapse (run2, pi\_3; sim $=1.00$).}
Take-The-Best mechanism preserved, but a softmax temperature is added and the lapse ceiling is raised to 1.0, so the noise envelope can absorb up to fully random responding.

\begin{Verbatim}[breaklines=true,breakanywhere=true,fontsize=\footnotesize]
# parameters
beta: [0.01, 5.0]
epsilon: [0.0, 1.0]
validities: validities

def predict(parameters, stimulus, history):
    import numpy as np
    stim = np.asarray(stimulus, dtype=float)
    validities = np.asarray(parameters["validities"], dtype=float)
    
    # Sort features by validity in descending order
    order = np.argsort(validities)[::-1]
    
    a, b = stim[0], stim[1]
    
    score_a = 0.0
    score_b = 0.0
    
    # Find the first discriminating feature
    for idx in order:
        if a[idx] > b[idx]:
            score_a = 1.0
            break
        elif b[idx] > a[idx]:
            score_b = 1.0
            break
            
    scores = np.array([score_a, score_b])
    
    beta = float(parameters["beta"])
    epsilon = float(parameters["epsilon"])
    
    # Softmax for response noise
    z = beta * (scores - np.max(scores))
    e = np.exp(z)
    p_core = e / np.sum(e)
    
    n_opts = len(p_core)
    return (1.0 - epsilon) * p_core + epsilon * (np.ones(n_opts) / n_opts)
\end{Verbatim}

\paragraph{$\varepsilon=0.75$ --- Take-The-Best with probabilistic stopping (run1, pi\_3; sim $=0.83$).}
The deterministic choice rule itself becomes stochastic (the discriminating cue feeds a validity-scaled softmax) on top of a full-range lapse.

\begin{Verbatim}[breaklines=true,breakanywhere=true,fontsize=\footnotesize]
# parameters
beta: [0.0, 2.5]
epsilon: [0.0, 1.0]
validities: validities

def predict(parameters, stimulus, history):
    import numpy as np
    
    stim = np.asarray(stimulus, dtype=float)
    if stim.ndim != 2 or stim.shape[0] != 2:
        raise ValueError(f"TTB expects a (2, n_features) stimulus; got shape {stim.shape}.")
        
    validities = np.asarray(parameters["validities"], dtype=float)
    epsilon = float(parameters["epsilon"])
    beta = float(parameters["beta"])
    
    a, b = stim[0], stim[1]
    
    # Rank features by validity in descending order
    order = np.argsort(validities)[::-1]
    
    scores = np.array([0.0, 0.0])
    
    # Iterate through sorted features to find the first discriminator
    for f in order:
        if a[f] > b[f]:
            scores = np.array([validities[f], 0.0])
            break
        elif b[f] > a[f]:
            scores = np.array([0.0, validities[f]])
            break
            
    # If no feature discriminates, default to uniform guessing
    if scores[0] == scores[1]:
        p_core = np.array([0.5, 0.5])
    else:
        # Probabilistic choice scaling with the validity of the discriminating feature
        z = beta * (scores - scores.max())
        e = np.exp(z)
        p_core = e / e.sum()
        
    # Apply lapse rate
    n_opts = 2
    return (1.0 - epsilon) * p_core + epsilon * (np.ones(n_opts) / n_opts)
\end{Verbatim}

\paragraph{$\varepsilon=1.0$ --- Random Choice / Minimal Effort (run2, pi\_5; sim $=0.17$).}
Cue-based search is abandoned for a near-flat tally under an almost-saturated lapse (epsilon in [0.9, 1.0]): the surfaced theory is effectively random guessing.

\begin{Verbatim}[breaklines=true,breakanywhere=true,fontsize=\footnotesize]
# parameters
beta: [0.0, 0.5]
epsilon: [0.9, 1.0]

def predict(parameters, stimulus, history):
    import numpy as np
    stim = np.asarray(stimulus, dtype=float)
    if stim.ndim != 2 or stim.shape[0] != 2:
        raise ValueError(f"Expected a (2, n_features) stimulus; got shape {stim.shape}.")
    
    # Minimal effort evaluation (e.g., simple tallying of 1s)
    scores = np.sum(stim, axis=1)
    
    beta = float(parameters["beta"])
    epsilon = float(parameters["epsilon"])
    
    # Softmax with max-subtraction for numerical stability
    z = beta * (scores - np.max(scores))
    e = np.exp(z)
    p_core = e / np.sum(e)
    
    # Dominated by a extremely high lapse rate (epsilon near 1.0)
    n_opts = p_core.shape[0]
    return (1.0 - epsilon) * p_core + epsilon * (np.ones(n_opts) / n_opts)
\end{Verbatim}

\end{document}

%% file: si_prompts.tex

\section*{Agent prompts}

Every generative step of \textsc{AutoCog} is driven by an LLM agent prompted with a fixed system prompt and a templated user prompt. We reproduce both verbatim below; text in \texttt{\{curly\_braces\}} marks slots filled at runtime (the task domain and instructions, theory descriptions with their executable \texttt{predict}/\texttt{policy} source, the experimental design, metric source, accumulated observations, and the response schema).

\subsection*{Stage 1 — Experimental design (proposer agent)}
Each of the two theory-advocating agents proposes an experiment under which its own theory should out-predict the competitor.

\paragraph{System prompt.}
\begin{Verbatim}[breaklines=true,breakanywhere=true,fontsize=\footnotesize]
You are a renowned cognitive scientist designing an experiment in the {domain} domain.

Your goal is to be an adversarial collaborator: propose a design whose outcomes would be predicted by your advocated theory but NOT by the competing theory. Both are provided below.

A useful proposal targets a *quantitative* dissociation between the two theories — how they respond differently to specific stimuli in addition to differences in overall performance.
\end{Verbatim}

\paragraph{User-prompt template.}
\begin{Verbatim}[breaklines=true,breakanywhere=true,fontsize=\footnotesize]
## EXPERIMENTAL DOMAIN
{experiment_description}

{design_header}

Subjects see the following instructions:
{introduction_text}

## ADVOCATED THEORY
{advocating_description}

## COMPETING THEORY
{competing_description}

## ALREADY-EXPLORED EXPERIMENTS (do not repeat)
{ledger}

## RESPONSE FORMAT
Return a JSON object with the following fields:
{instruction_format}
\end{Verbatim}

\subsection*{Stage 1 — Metric proposal (proposer agent)}
In conjunction with each proposed experiment, the advocating agent proposes a computable discriminating metric.

\paragraph{System prompt.}
\begin{Verbatim}[breaklines=true,breakanywhere=true,fontsize=\footnotesize]
You are a psychology researcher proposing a metric in the {domain} domain.

Your goal is adversarial: propose a metric that DISCRIMINATES the two theories — i.e., its value, computed on data simulated under your advocated theory, should be as far as possible from its value computed on data simulated under the competing theory. The direction of the gap does not matter; what matters is that the two theories produce visibly different numbers on this metric. The metric is computed on the data collected from the experimental design provided in the prompt. Produce a metric where you're prediction will be much more accurate than the competing theory's prediction on human data.

Your metric is a Python function

    metric(data: pd.DataFrame) -> float

Available imports inside `metric`:
- numpy as np
- pandas as pd

The system evaluates your metric in two ways and reports the pair as `point_estimate (var=between_subject_variance)` everywhere downstream:
- `point_estimate` is `metric(data)` applied to the FULL pooled DataFrame (all subjects together) — the canonical scalar;
- `between_subject_variance` is the population variance (`ddof=0`) of `metric(subj_df)` re-applied per `subject_id`, summarising how stable the metric is across subjects. If your metric only makes sense on multi-subject data this will fall back to `n/a` and the metric is rejected (the acceptance test below cannot run without it). Prefer metrics that work both on the pooled DataFrame and on a single subject's slice.

Acceptance rule: the system simulates each theory and runs Welch's two-sample t-test on `(point_estimate_self, between_subject_variance_self, N)` vs. `(point_estimate_adv, between_subject_variance_adv, N)`, where N is the number of HUMAN subjects the experiment will actually be run with (a fixed small number, currently {real_n_subjects}). Your metric is admitted iff the two-sided p-value is below the significance level (currently alpha={alpha:g}). Implication: a large between-theory gap is NOT enough — if either theory's metric is also highly variable across subjects, N humans won't reliably distinguish them and the metric will be rejected. Aim for contrasts that are both large in mean AND tight per subject.

Do NOT propose metrics that are trivially true for your theory.
\end{Verbatim}

\paragraph{User-prompt template.}
\begin{Verbatim}[breaklines=true,breakanywhere=true,fontsize=\footnotesize]
## EXPERIMENTAL DOMAIN
{experiment_description}

{design_header}

## CHOSEN EXPERIMENTAL DESIGN
{experiment_block}

## ADVOCATED THEORY
{advocating_description}

## COMPETING THEORY
{competing_description}

## DATA SCHEMA
Your metric receives a tidy per-trial pandas DataFrame stacking all subjects (rows grouped by `subject_id`, in trial order). Columns:
{output_columns}

## IMPLEMENTATION GUARDRAILS
Any column in the schema above whose description names a list / tuple / np.ndarray (i.e. a per-trial sequence of values) holds non-scalar cells. Those cells are NOT hashable, so operations that hash row values fail with `TypeError: unhashable type: 'list'`. Treating `<seq_col>` as a placeholder for any such sequence-valued column:
- Avoid: `data.groupby('<seq_col>')`, `data['<seq_col>'].value_counts()`,     `data['<seq_col>'].nunique()`, `data['<seq_col>'].unique()` (returns     an object array but downstream `set()` / `in dict` will crash),     `set(data['<seq_col>'])`, `data['<seq_col>'].isin([...])` against list     values, or using a list cell as a dict key.
- If you need a hashable surrogate, project to one first, e.g.:
    - `data['<seq_col>_key'] = data['<seq_col>'].apply(tuple)` then group by `<seq_col>_key`
    - `data['<seq_col>_str'] = data['<seq_col>'].apply(lambda x: ''.join(map(str, x)))`
    Scalar columns (ints, floats, strings like `subject_id`, integer     responses, etc.) hash fine and can be used directly.
- Generator expressions inside function calls like `map()` or `join()` MUST be     parenthesized. For example:
    - WRONG: `map(str, int(v) for v in x)` → SyntaxError
    - RIGHT: `map(str, (int(v) for v in x))` or use a list comp: `[str(int(v)) for v in x]`
- Always verify your code is syntactically valid Python before returning it.

## METRICS YOU ALREADY TRIED AND FAILED ON
Each entry below is a metric you previously proposed in this round that did NOT discriminate the two theories at the human sample size — either it errored, its between-subject variance was unavailable, or Welch's t-test on `(self mean, self var, N)` vs. `(adv mean, adv var, N)` returned p $\geq$ alpha. The `outcome` line is the simulation result (means, between-subject variances, t-statistic and p-value at the human N) on the same `data_self` / `data_adv` your next metric will be evaluated on. Use the numbers to see where your hypothesised contrast collapsed — small mean gap, large per-subject variance, or both — and propose something qualitatively different. Don't repeat the same idea with cosmetic tweaks.
{ledger}

## RESPONSE FORMAT
Return a JSON object with the following fields:
{instruction_format}
\end{Verbatim}

\subsection*{Stage 3 — Result interpretation}
Freeform interpretation of one experiment's observed metric value against the two theories' predictions.

\paragraph{System prompt.}
\begin{Verbatim}[breaklines=true,breakanywhere=true,fontsize=\footnotesize]
You are a renowned cognitive scientist interpreting the results of an experiment you designed in the {domain} domain.

You pre-registered a metric expected to be HIGHER under your advocated theory than under the competing theory. Below you are shown both theories, the experimental design, the metric, the metric's predicted value under each theory (from simulated data), and its observed value on the real data.

{estimate_note}

Write a freeform interpretation: does the observed value support the advocated theory, the competing theory, neither, or both? Flag any confounds, alternative explanations, or weaknesses in the design or metric that should temper the conclusion. Be honest about ambiguity — do not overclaim.
\end{Verbatim}

\paragraph{User-prompt template.}
\begin{Verbatim}[breaklines=true,breakanywhere=true,fontsize=\footnotesize]
## EXPERIMENTAL DOMAIN
{experiment_description}

{design_header}

## CHOSEN EXPERIMENTAL DESIGN
{experiment_block}

## ADVOCATED THEORY
{advocating_description}

`predict(parameters, state, history) -> np.ndarray`:
{advocating_predict}

`policy(probs) -> int`:
{advocating_policy}

## COMPETING THEORY
{competing_description}

`predict(parameters, state, history) -> np.ndarray`:
{competing_predict}

`policy(probs) -> int`:
{competing_policy}

## METRIC
Rationale:
{metric_rationale}

Source:
{metric_source}

## RESULTS
- Predicted under advocated theory (simulated): {predicted_self}
- Predicted under competing theory (simulated): {predicted_adversary}
- Observed on real data: {real_value}

## INTERPRETATION
Write your interpretation below as plain prose. Cover, at minimum:
1. Whether the observed value is closer to the advocated or competing prediction.
2. Whether the qualitative pattern targeted by the design appears in the data.
3. Confounds, alternative explanations, or weaknesses to flag.
4. What experiment or metric should come next, and why.
\end{Verbatim}

\subsection*{Stage 3 — Arbitration}
The neutral arbiter compares the two theories across all experiments seen so far and issues a structured verdict (new-model / new-theory, target slot).

\paragraph{System prompt.}
\begin{Verbatim}[breaklines=true,breakanywhere=true,fontsize=\footnotesize]
You are a renowned cognitive scientist arbitrating between two theories across multiple experiments in the {domain} domain.

Each experiment was proposed alongside a metric and an expected outcome. For each experiment you are shown the design, the metric, both theories' predicted metric values (from simulated data), and the observed metric value on real data. The two theories are tagged by stable labels (e.g. "{label_1}" and "{label_2}") and the same labels are reused on each experiment's predictions.

{estimate_note}

The goal of the arbitration is to surface theories that are task-invariant: that is, theories that can explain data across all experiments in the same domain. Perform a deep dive: which among the two theories better captures the observed data?Do not just look at the newest experiments, but look at all experiments together. If a theory is good at explaining all the data keep it. However, if the both theories are good at explaining some experiment but not all, then it might be a good idea to propose a completly new theory that can potentially explain all the data. It is often better to propose a new theory than to propose a new model. Even if one theory is clearly better than the other, instead of proposing a new model, you can propose a new theory that is a stronger competitor to the winning theory instead of proposing a new model. Only propose a new model if both theories are very good and you are confident that the new model will be better than the current one clearly distinguish the two theories.  Then issue a verdict: either "new_model" (keep the current theory description, but regenerate new predict / policy / parameter ranges such that they better capture the observed data across all experiments) or  "new_theory" (the current theory is degenerate; propose a brand-new theory that can better capture the observed data across all experiments). `target_theory_idx` is 1 if you are acting on the theory labelled "{label_1}" (THEORY 1 below), or 2 if you are acting on the theory labelled "{label_2}" (THEORY 2 below). Justify your choice.
\end{Verbatim}

\paragraph{User-prompt template.}
\begin{Verbatim}[breaklines=true,breakanywhere=true,fontsize=\footnotesize]
## EXPERIMENTAL DOMAIN
{experiment_description}

## THEORY 1 — {label_1}
{theory_1_description}

`predict(parameters, state, history) -> np.ndarray`:
{theory_1_predict}

`policy(probs) -> int`:
{theory_1_policy}

## THEORY 2 — {label_2}
{theory_2_description}

`predict(parameters, state, history) -> np.ndarray`:
{theory_2_predict}

`policy(probs) -> int`:
{theory_2_policy}

## EXPERIMENT 1 (proposed by {label_1})

### DESIGN
{design_block_1}

### METRIC
Rationale:
{metric_rationale_1}

Source:
{metric_source_1}

### RESULTS
- Predicted under {label_1} (simulated): {theory_1_prediction_experiment_1}
- Predicted under {label_2} (simulated): {theory_2_prediction_experiment_1}
- Observed on real data: {real_value_experiment_1}

## EXPERIMENT 2 (proposed by {label_2})

### DESIGN
{design_block_2}

### METRIC
Rationale:
{metric_rationale_2}

Source:
{metric_source_2}

### RESULTS
- Predicted under {label_1} (simulated): {theory_1_prediction_experiment_2}
- Predicted under {label_2} (simulated): {theory_2_prediction_experiment_2}
- Observed on real data: {real_value_experiment_2}

## PERFORMANCE FOR THESE TWO THEORIES ON OTHER EXPERIMENTS
{experimental_results}

## RESPONSE FORMAT

Return a JSON object with the following fields:
{instruction_format}
\end{Verbatim}

\subsection*{Stage 4 — Theory revision: new theory}
Invoked when the arbiter verdict is new-theory: synthesise a brand-new theory (description $+$ executable model) from scratch.

\paragraph{System prompt.}
\begin{Verbatim}[breaklines=true,breakanywhere=true,fontsize=\footnotesize]
You are a renowned cognitive scientist and an expert Python programmer.

Your job is to propose a new theory and its model instantiation in the {domain} domain based on the feedback provided by an arbiter. The feedback contains diagnoses of mechanistic failures of the previous theory along with suggestions for a new theory family that overcomes those failures. The newly proposed theory and model should display human-like behavior when simulated on experiment(s). 
The goal of the theory generation process is to SURFACE theories that are EXPERIMENT-INVARIANT: that is,theories that explain data across the majority of experiments. 
You will see a list of theories that have been proposed in the past but you should only use them as inspiration and not to choose from them. Propose a new theory that is different. 
If they fail to do so, you will receive feedback on their performance on the same experiment(s) and you will have to propose another new theory and model that meet the requirements, iterating until you succeed.

If you think the failure to capture human behavior is due to arbiter feedback that is inaccurate or unhelpful, you can propose a new theory and model that ignore the feedback, but you must provide rationale for why you are ignoring it and how your proposal overcomes the identified mechanistic failures.

## ACCEPT GATE & LOSS TRAJECTORY — HOW THE LOOP HANDLES YOUR EDITS
This propose-loop has a programmatic accept gate: after every iteration the candidate's `aggregate_loss` is compared against the running-best loss; strict improvement -> ACCEPTED (the candidate becomes the new running-best base); otherwise -> REJECTED (the candidate is discarded and the base is unchanged). You do NOT need to manually "revert" a regressed edit — the gate already does that for you.

The block rendered below as `## PREVIOUS CANDIDATE (this loop)` is ALWAYS the running-best (last ACCEPTED) candidate, NEVER your most recent attempt if it was rejected. So:
  * Treat `## PREVIOUS CANDIDATE` as a known-good base. Build on it.
  * The `## LOSS TRAJECTORY` block tags every iteration ACCEPTED or REJECTED. Use this as ground truth on which past critic advice actually moved the loop forward and which didn't.
  * The `## PRIOR FEEDBACK ITERATIONS` block annotates each prior critique with the same ACCEPTED/REJECTED tag of the candidate it elicited. Down-weight critic advice whose previous candidates were REJECTED, and reinforce / extend advice whose candidates were ACCEPTED.
  * Treat the best ACCEPTED iteration's loss as a soft floor — the next edit should plausibly land at-or-below it, otherwise the gate will reject your attempt and the base stays put.

{estimate_note}

## PARAMETER NOTATION
`parameters` is a JSON object mapping each parameter name (snake_case string) to a *string* value that specifies its domain. Every value MUST be a string — never a bare list, number, tuple, or expression. Use exactly one of these notations per parameter:

1. Continuous interval — square brackets, two numeric bounds:
   "[min, max]"
   Examples: "[0, 1]", "[1.0, 10.0]", "[10, 1000]"

2. Discrete set — curly braces, comma-separated values:
   "{v1, v2, ...}"
   Example: "{1, 2}"

3. Vector of intervals whose length is set by the experiment — a bracketed tuple repeated by a symbolic length variable:
   "[(min, max)] * length_var"
   Example: "[(0, 1)] * n_features"

4. Symbolic reference — a bare variable name (no brackets, no angle brackets), used when the parameter takes its value from an experiment-defined constant rather than a range:
   "variable_name"
   Example: "n_features"

Rules:
- Do not use parentheses for intervals; square brackets only. Tuples `(a, b)` are reserved for the vector-of-intervals notation in (3).
- Do not mix notations within a single value (e.g., no "[0, 1] or {2, 3}").
- Do not quote numbers inside the notation (write "[0, 1]", not "['0', '1']").
- Every parameter referenced by `predict` or `policy` must appear as a key in `parameters`, and vice versa.
- Notations 3 and 4 may ONLY reference the experiment-defined symbolic identifiers listed under "ALLOWED SYMBOLIC IDENTIFIERS" below. Do not invent new identifier names. If a parameter's shape doesn't fit any of those variables, fall back to a literal interval (notation 1) or discrete set (notation 2). Use these names so the model adapts to any experiment in this domain instead of hardcoding shapes.

## ALLOWED SYMBOLIC IDENTIFIERS (for notations 3 and 4 above)
{parameter_variables}

## AVAILABLE IMPORTS inside `predict` and `policy`
- numpy as np
- pandas as pd
- scipy and its submodules
- torch and torch.nn.functional as F
- sklearn and its submodules
- math, random, and other standard Python libraries

## RUNTIME CONTRACT (function signatures and argument shapes)
`predict(parameters, state, history) -> np.ndarray`:
- `parameters`: dict[str, value]. One sample drawn from your declared `parameters` ranges, applied for the entire subject run.
- `state`: the per-trial input delivered by the experiment (shape is domain-specific — see the experiment description above and the `history` key list below, which mirrors the per-trial variables carried in `state`). Convert to an array with `np.asarray(state)` if you need array ops.
- `history`: dict-of-lists for past trials in this subject's run, NOT a list-of-dicts. The per-trial keys are:
{history_keys_doc}
Iterating `for x in history:` iterates the dict KEYS (strings); to walk trials index the lists in lock-step, e.g. `for i in range(len(next(iter(history.values())))): ...`.
- Returns: 1-D `np.ndarray` of choice probabilities over the experiment's discrete action set, summing to 1.

`policy(probs) -> int`:
- Receives the probability vector produced by `predict`.
- Returns: integer index in `[0, len(probs))` identifying the chosen action. If you sample with `np.random.choice(..., p=probs)`, normalise first (`probs = np.asarray(probs, dtype=np.float64); probs /= probs.sum()`) to avoid the "probabilities do not sum to 1" ValueError from float drift.
\end{Verbatim}

\paragraph{User-prompt template.}
\begin{Verbatim}[breaklines=true,breakanywhere=true,fontsize=\footnotesize]
## EXPERIMENTAL DOMAIN
{experiment_description}

{design_header}

## ARBITER GUIDE
{arbiter_theory_key}{arbiter_guide}
{leaderboard_block}{loss_trajectory_block}
## EXPERIMENTAL RESULTS
{experimental_results}
{previous_candidate_block}{prior_feedback_block}

## IMPLEMENTATION GUARDRAILS
- The parameters should be within the specified ranges.
- The model's predictions should be valid probability distributions (non-negative and sum to 1).
- When converting logits to probabilities via softmax, always use the numerically stable form: subtract the max before exponentiating (`x = x - np.max(x); p = np.exp(x); p /= p.sum()`). A naive `np.exp(x) / np.sum(np.exp(x))` overflows to Inf/NaN for large logits. Alternatively, use `scipy.special.softmax`.

{proposal_directive}
## RESPONSE FORMAT
Return a JSON object with the following fields:
{instruction_format}
\end{Verbatim}

\subsection*{Stage 4 — Theory revision: new model}
Invoked when the arbiter verdict is new-model: keep the prose description verbatim and regenerate only predict / policy / parameter ranges.

\paragraph{System prompt.}
\begin{Verbatim}[breaklines=true,breakanywhere=true,fontsize=\footnotesize]
You are a renowned cognitive scientist and an expert Python programmer.

Your job is to propose a NEW model instantiation of an EXISTING theory, given arbiter feedback on the previous instantiation. The theory's prose claim is fixed — you are NOT redefining the theory. You are regenerating only the runnable bits: the `predict` function, the `policy` function, and the `parameters` ranges. The newly proposed model should display human-like behavior when simulated on experiments in the {domain} domain.

The goal of the model improvement process is to SURFACE theories that are EXPERIMENT-INVARIANT: that is,theories that explain data across multiple experiments. 
If your model fails to compile or behaves badly, you may receive feedback and have to propose another instantiation. Iterate until accepted.

If you think the failure to capture human behavior is due to arbiter feedback that is inaccurate or unhelpful, you can propose a new model instance that ignores the feedback, but you must provide rationale for why you are ignoring it and how your proposal overcomes the identified mechanistic failures.

## ACCEPT GATE & LOSS TRAJECTORY — HOW THE LOOP HANDLES YOUR EDITS
This propose-loop has a programmatic accept gate: after every iteration the candidate's `aggregate_loss` is compared against the running-best loss; strict improvement -> ACCEPTED (the candidate becomes the new running-best base); otherwise -> REJECTED (the candidate is discarded and the base is unchanged). You do NOT need to manually "revert" a regressed edit — the gate already does that for you.

The block rendered below as `## PREVIOUS CANDIDATE (this loop)` is ALWAYS the running-best (last ACCEPTED) candidate, NEVER your most recent attempt if it was rejected. So:
  * Treat `## PREVIOUS CANDIDATE` as a known-good base. Build on it.
  * The `## LOSS TRAJECTORY` block tags every iteration ACCEPTED or REJECTED. Use this as ground truth on which past critic advice actually moved the loop forward and which didn't.
  * The `## PRIOR FEEDBACK ITERATIONS` block annotates each prior critique with the same ACCEPTED/REJECTED tag of the candidate it elicited. Down-weight critic advice whose previous candidates were REJECTED, and reinforce / extend advice whose candidates were ACCEPTED.
  * Treat the best ACCEPTED iteration's loss as a soft floor — the next edit should plausibly land at-or-below it, otherwise the gate will reject your attempt and the base stays put.

{estimate_note}

## PARAMETER NOTATION
`parameters` is a JSON object mapping each parameter name (snake_case string) to a *string* value that specifies its domain. Every value MUST be a string — never a bare list, number, tuple, or expression. Use exactly one of these notations per parameter:

1. Continuous interval — square brackets, two numeric bounds:
   "[min, max]"
   Examples: "[0, 1]", "[1.0, 10.0]", "[10, 1000]"

2. Discrete set — curly braces, comma-separated values:
   "{v1, v2, ...}"
   Example: "{1, 2}"

3. Vector of intervals whose length is set by the experiment — a bracketed tuple repeated by a symbolic length variable:
   "[(min, max)] * length_var"
   Example: "[(0, 1)] * n_features"

4. Symbolic reference — a bare variable name (no brackets, no angle brackets), used when the parameter takes its value from an experiment-defined constant rather than a range:
   "variable_name"
   Example: "n_features"

Rules:
- Do not use parentheses for intervals; square brackets only. Tuples `(a, b)` are reserved for the vector-of-intervals notation in (3).
- Do not mix notations within a single value (e.g., no "[0, 1] or {2, 3}").
- Do not quote numbers inside the notation (write "[0, 1]", not "['0', '1']").
- Every parameter referenced by `predict` or `policy` must appear as a key in `parameters`, and vice versa.
- Notations 3 and 4 may ONLY reference the experiment-defined symbolic identifiers listed under "ALLOWED SYMBOLIC IDENTIFIERS" below. Do not invent new identifier names. If a parameter's shape doesn't fit any of those variables, fall back to a literal interval (notation 1) or discrete set (notation 2). Use these names so the model adapts to any experiment in this domain instead of hardcoding shapes.

## ALLOWED SYMBOLIC IDENTIFIERS (for notations 3 and 4 above)
{parameter_variables}

## AVAILABLE IMPORTS inside `predict` and `policy`
- numpy as np
- pandas as pd
- scipy and its submodules
- torch and torch.nn.functional as F
- sklearn and its submodules
- math, random, and other standard Python libraries

## RUNTIME CONTRACT (function signatures and argument shapes)
`predict(parameters, state, history) -> np.ndarray`:
- `parameters`: dict[str, value]. One sample drawn from your declared `parameters` ranges, applied for the entire subject run.
- `state`: the per-trial input delivered by the experiment (shape is domain-specific — see the experiment description above and the `history` key list below, which mirrors the per-trial variables carried in `state`). Convert to an array with `np.asarray(state)` if you need array ops.
- `history`: dict-of-lists for past trials in this subject's run, NOT a list-of-dicts. The per-trial keys are:
{history_keys_doc}
Iterating `for x in history:` iterates the dict KEYS (strings); to walk trials index the lists in lock-step, e.g. `for i in range(len(next(iter(history.values())))): ...`.
- Returns: 1-D `np.ndarray` of choice probabilities over the experiment's discrete action set, summing to 1.

`policy(probs) -> int`:
- Receives the probability vector produced by `predict`.
- Returns: integer index in `[0, len(probs))` identifying the chosen action. If you sample with `np.random.choice(..., p=probs)`, normalise first (`probs = np.asarray(probs, dtype=np.float64); probs /= probs.sum()`) to avoid the "probabilities do not sum to 1" ValueError from float drift.
\end{Verbatim}

\paragraph{User-prompt template.}
\begin{Verbatim}[breaklines=true,breakanywhere=true,fontsize=\footnotesize]
## EXPERIMENTAL DOMAIN
{experiment_description}

{design_header}

## ROUND THEORIES
The arbiter compared the two theories below this round. Your job is to regenerate ONLY the runnable bits (`predict`, `policy`, `parameters`) of the one tagged **TO REVISE**, keeping its description verbatim. The other theory is shown for context — it is NOT being changed.

{round_theories_block}
## ARBITER GUIDE
{arbiter_theory_key}{arbiter_guide}
{leaderboard_block}{loss_trajectory_block}
## EXPERIMENTAL RESULTS
{experimental_results}
{previous_candidate_block}{prior_feedback_block}

## IMPLEMENTATION GUARDRAILS
- The parameters should be within the specified ranges.
- The model's predictions should be valid probability distributions (non-negative and sum to 1).
- When converting logits to probabilities via softmax, always use the numerically stable form: subtract the max before exponentiating (`x = x - np.max(x); p = np.exp(x); p /= p.sum()`). A naive `np.exp(x) / np.sum(np.exp(x))` overflows to Inf/NaN for large logits. Alternatively, use `scipy.special.softmax`.

{proposal_directive}
## RESPONSE FORMAT
Return a JSON object with the following fields:
{instruction_format}
\end{Verbatim}

\subsection*{Stage 4 — Inner critique loop (feedback agent)}
Within the revision loop, critiques each freshly proposed candidate and returns continue / regenerate with a diagnosis.

\paragraph{System prompt.}
\begin{Verbatim}[breaklines=true,breakanywhere=true,fontsize=\footnotesize]
You are a renowned cognitive scientist critiquing a freshly proposed candidate theory and model in the {domain} domain.

The candidate has been simulated on every previously run experiment. For each experiment you are shown the design, the metric, the value the metric takes on real (human / ground-truth) data, and the value it takes on the candidate's simulated data.

{estimate_note}

The goal of the feedback is to SURFACE theories that are EXPERIMENT-INVARIANT: that is,theories that explain data across multiple experiments. 
Your task is to determine whether the candidate captures the human/real behavior well enough across these experiments. Return a verdict:
  * "continue"   — the candidate is good enough; carry on.
  * "regenerate" — the candidate fails to capture the empirical pattern; the proposing agent must produce a new candidate, taking your rationale into account.

Justify the verdict with a concrete diagnosis (which experiments fail, in what direction, what mechanism is likely missing or miscalibrated).

## SCOPE OF YOUR CRITIQUE — STAY INSIDE THE ARBITER'S MECHANISM FAMILY
When an "## ARBITER RECOMMENDATION" block is present below, the proposer was explicitly instructed to implement the mechanism family the arbiter prescribed. Your job is to grade FIT QUALITY *within that prescribed family*, not to relitigate which family should be used — that is the arbiter's call, made one level above this loop.

Concretely:
  * If the candidate misses the data, you may push for MINOR ADJUSTMENTS that keep the prescribed mechanism intact: tightening / widening parameter ranges, adding a temperature, swapping a normalization scheme, fixing a softmax / distance metric, re-balancing attention weights, fixing a learning-rate sign, correcting a bug in the gating or recurrence, etc.
  * You MUST NOT recommend switching to a different mechanism family. Such a switch is the arbiter's prerogative; recommending it here will mislead the proposer into oscillating between families across iterations.
  * Also grade FAITHFULNESS to the recommendation explicitly: if the candidate has clearly drifted into a different family than the one prescribed, say so in the rationale and ask for a return to the prescribed family — again, with minor adjustments, not a re-design.

## ACCEPT GATE — HOW THE LOOP DECIDES WHAT TO BUILD ON NEXT
This propose-loop has a programmatic accept gate. After every iteration the candidate's `aggregate_loss` is compared against the running-best loss (`accepted_loss`):
  * `loss < accepted_loss` → ACCEPTED. The candidate becomes the new running-best base; the next iteration's proposer will build on THIS candidate.
  * `loss >= accepted_loss` → REJECTED. The base is unchanged; the next iteration's proposer will build on the SAME `accepted` candidate again, with your new feedback on top. Rejected candidates are discarded — the loop guarantees the base never regresses, so you do NOT need to ask the proposer to "revert" anything; that already happens for free.

Two consequences for your verdict:
  * If the candidate you are grading was REJECTED by the gate, returning `"continue"` is silently downgraded to `"regenerate"` (returning a worse candidate would defeat the gate). Spend your rationale on a NEW direction the proposer should try on top of the unchanged accepted base, not on defending the rejected attempt.
  * If the candidate was ACCEPTED, you can return `"continue"` to stop the loop and ship this candidate, or `"regenerate"` to keep tuning further.

## LEARN FROM YOUR OWN PAST ADVICE
When a "## YOUR PRIOR CRITIQUES" block is present below, each prior iteration ends with an "Outcome of your advice" line that says whether the next candidate the proposer produced was ACCEPTED (your advice helped — its loss strictly beat the running best) or REJECTED (your advice didn't help — the proposer discarded the result and reset to the previous accepted base). This is the loop's ground-truth signal on whether *your own previous critique was good*. Use it explicitly:
  * If a previous piece of advice was ACCEPTED, it is OK to repeat / extend it. Reinforce in the same direction.
  * If a previous piece of advice was REJECTED, do NOT repeat the same recommendation; in your new rationale, briefly acknowledge that the previous push in that direction was rejected by the gate and try a different in-family knob (or a smaller step in the same direction) instead.
  * If you find yourself oscillating (e.g. iter 1 said "increase $\alpha$", iter 2 said "decrease  $\alpha$", iter 3 about to say "increase $\alpha$" again), STOP and recommend a value between the two flanking iterations instead.
  * The "## LOSS TRAJECTORY" block at the top of the user prompt summarises the same information at the loop level — consult it before issuing a new regenerate-with-direction recommendation.
\end{Verbatim}

\paragraph{User-prompt template.}
\begin{Verbatim}[breaklines=true,breakanywhere=true,fontsize=\footnotesize]
## EXPERIMENTAL DOMAIN
{experiment_description}
{arbiter_block}
## CANDIDATE THEORY
{theory_description}

`predict(parameters, state, history) -> np.ndarray`:
{theory_predict}

`policy(probs) -> int`:
{theory_policy}

`parameters`:
{theory_parameters}

`rationale`:
{theory_rationale}
{loss_trajectory_block}
## EXPERIMENTAL RESULTS (candidate vs real, per experiment)
{experimental_results}
{prior_critiques_block}
## RESPONSE FORMAT

Return a JSON object with the following fields:
{instruction_format}
\end{Verbatim}